\documentclass[journal=jacsat,manuscript=article]{achemso}
\usepackage[version=3]{mhchem} % Formula subscripts using \ce{}
\usepackage[dvipsnames]{xcolor}

\usepackage[normalem]{ulem} % allow to bar text with \sout{}
\definecolor{blue}{rgb}{0,0,1}
\definecolor{cyan}{rgb}{0.2,0.7,1}

\author{Vasily Artemov}
\affiliation{Laboratory of Nanoscale Biology, Institute of Bioengineering, École Polytechnique Fédérale de Lausanne (EPFL), 1015 Lausanne, Switzerland}
\email{vasily.artemov@epfl.ch}

\author{Laura Frank\textsuperscript{1}}
\affiliation{Institute of Meteorology and Climate Research, Karlsruhe Institute of Technology, 76021 Karlsruhe, Germany}

\author{Roman Doronin\textsuperscript{2}}
\affiliation{Institute of Meteorology and Climate Research, Karlsruhe Institute of Technology, 76021 Karlsruhe, Germany}

\author{Philipp St\"ark}
\affiliation{SC Simtech, University of Stuttgart, 70569 Stuttgart, Germany}

\author{Alexander Schlaich}
\affiliation{SC Simtech, University of Stuttgart, 70569 Stuttgart, Germany}
\alsoaffiliation{Institute for Computational Physics, University of Stuttgart, 70569 Stuttgart, Germany}

\author{Anton Andreev}
\affiliation{Department of Physics, University of Washington, 98195 Seattle, Washington, USA}

\author{Thomas Leisner}
\affiliation{Institute of Meteorology and Climate Research, Karlsruhe Institute of Technology, 76021 Karlsruhe, Germany}

\author{Aleksandra Radenovic}
\affiliation{Laboratory of Nanoscale Biology, Institute of Bioengineering, École Polytechnique Fédérale de Lausanne (EPFL), 1015 Lausanne, Switzerland}

\author{Alexei Kiselev}
\affiliation{Institute of Meteorology and Climate Research, Karlsruhe Institute of Technology, 76021 Karlsruhe, Germany}

\title[An \textsf{achemso} demo]
{Elucidating contact electrification mechanism of water}

\begin{document}

\begin{abstract}

The open water surface is known to be charged. Yet, the magnitude of the charge and the physical mechanism of the charging remain unclear, causing heated debates across the scientific community. Here we directly measure the charge $Q$ of microdrops ejected from hydrophilic and hydrophobic capillaries and show that the water surface can take both positive or negative charge values depending on pH and the capillary type. Our experiments, theory, and simulations provide evidence that a junction of two aqueous interfaces with a different ion adsorption energy (e.g., liquid-solid and liquid-air interfaces) develops a pH-dependent contact potential difference $\Delta\phi$ up to 52 mV. The longitudinal charge transfer between the interfaces stimulated by $\Delta\phi$ determines the charge of the open water surface. The suggested static electrification mechanism provides far-reaching insights into the origin of electrical potentials in biological and electrochemical energy systems.

\end{abstract}

\begin{figure}
\includegraphics{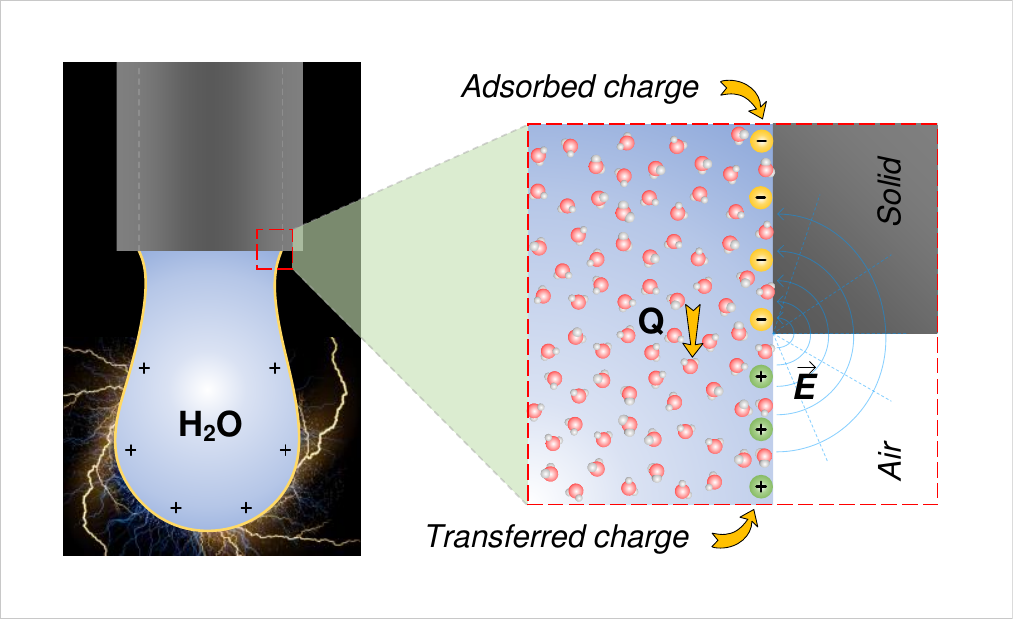}
\centering
\end{figure}

\footnotetext[1]{Currently at Steinbuch Centre for Computing, Karlsruhe Institute of Technology, 76128 Karlsruhe}
\footnotetext[2]{Currently at Datana LLC}

\section{Introduction}

Electrified aqueous interfaces appear in a myriad of natural phenomena and technological processes. The surface charging mechanism has important implications for atmospheric science~\cite{Saunders1991}, biological transport~\cite{Lowell1980}, nanofluidics ~\cite{Holt2006, Grosjean2019, Sun2021a}, static electrification~\cite{Thomson1867, Helseth2015}, electrochemical power systems~\cite{Stamenkovic2017, Dong2022, Ponomarenko2022}, and the understanding of fundamental properties of water~\cite{Artemov2021}. However, there is a puzzle: the information on the magnitude and the sign of the charge at the water interface is contradictory~\cite{Yatsuzuka1994,  Binks2001, Duft2003, Mora2007, Buch2007, Beattie2008, Petersen2008, Logan2008, Apodaca2010, Baytekin2011, Santos2011, Saykally2013, Miljkovic2013, Lin2014, Zhu2014, Donaldson2015, Hua2015, Shavlov2018, Wang2019, Shin2021, Pullanchery2021}. A molecular-level mechanism that drives the charging of water in each particular case lacks clarity~\cite{Chaplin2009, Sun2021}, though there is progress in this direction~\cite{Becker2022}. The situation hinders the development of modern electrochemical and energy-harvesting systems, as well as limits our understanding 
 of the mechanisms of natural electrification on the scales from microscopic to global. This unsolved interdisciplinary problem attracts scientists from different fields. Previous studies of the phenomenon of contact electrification revealed various processes, including the mass transfer across the liquid-air interface~\cite{Shavlov2018, Santos2011}, charging due to the separation of ions~\cite{Logan2008}, surface chemical reactions~\cite{Hua2015, Zhu2014}, fluctuations of surface tension~\cite{Apodaca2010}, electron transfer reactions~\cite{Wang2019} and the contribution from nano-molar impurities~\cite{uematsu2018}. Nevertheless, the primary mechanism of water surface electrification remains elusive~\cite{Chaplin2009, Sun2021}. 

Here, we report measurements of the charging of the water microdrops in a wide range of pH by ejecting them from capillaries of different types and varying the external electric potential. Our study, supported by theory and molecular dynamics simulations, provides evidence for a contact potential difference $\Delta\phi$ between two different (e.g., liquid-solid and liquid-air) interfaces, which is similar to that between different terminating crystal surfaces or between different metals~\cite{Landau8}. This voltage studied in semiconductor heterostructures~\cite{Meyerhof1947}, has never been discussed for the junctions of aqueous interfaces, even though the charge separation in an electrical double layer has been studied for a long time~\cite{deGennes1985, Derjaguin, Israelachvili2011}. Considering the difference in surface adsorption of ions on different surfaces, we show that a junction of two different aqueous interfaces develops a potential difference $\Delta\phi$ up to 52 mV. The surface with a lower adsorbed charge (e.g., open-water surface) charges by $\Delta\phi$ due to the charge transfer across the junction of different interfaces. This previously overlooked mechanism suggests a simple explanation of contact electrification of aqueous interfaces.

\section*{Experimental}

\begin{figure}
    \centering
    \includegraphics[width=\linewidth]{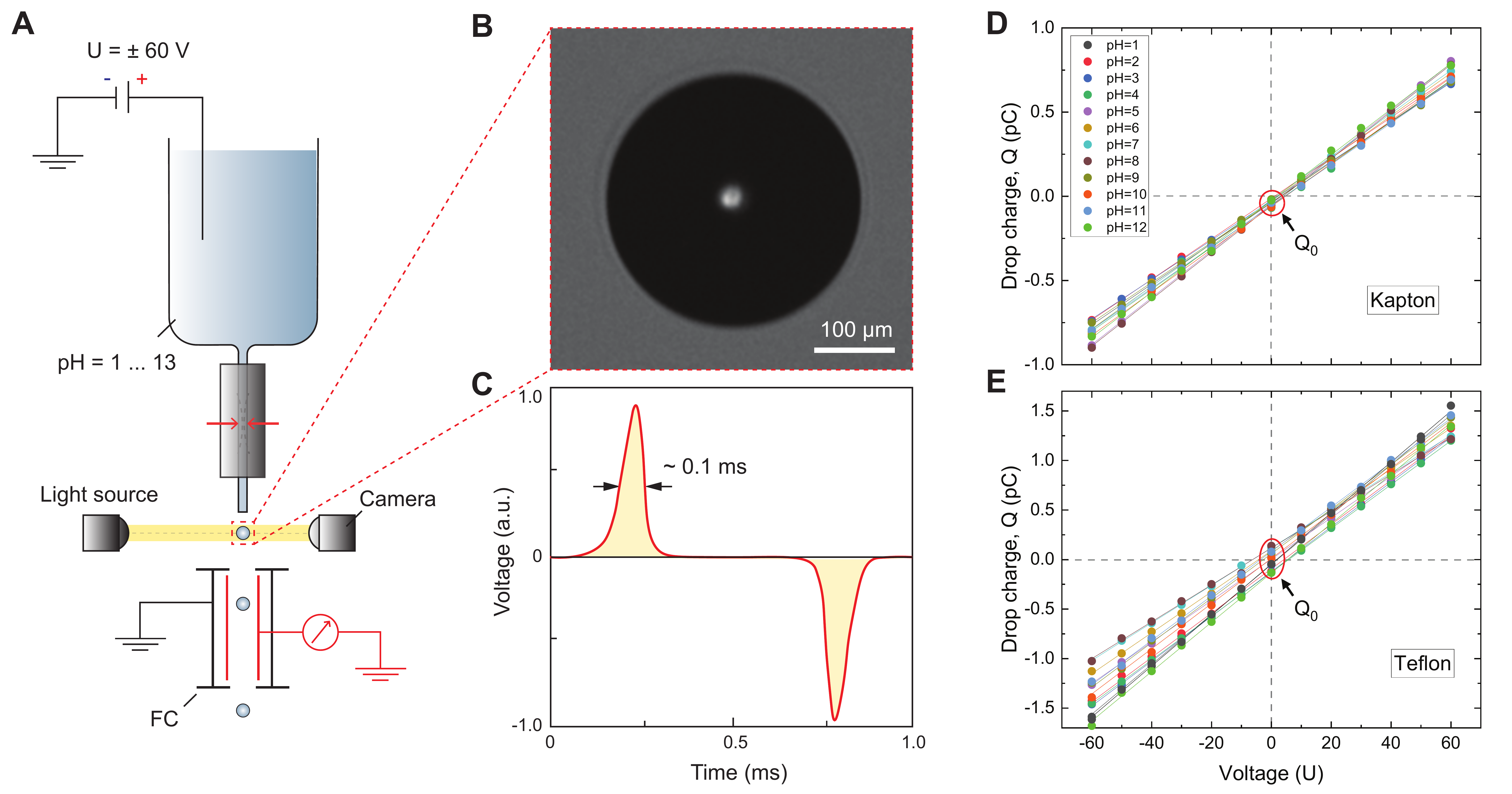}
    \caption{\textbf{Microdrop charge measurements.} (\textbf{A}) Setup generating monodisperse microdrops. The Faraday cylinder (FC) measures the voltage between the inner and outer shells induced by the falling microdrop. $U$ is the external voltage. The blue area is an aqueous solution with a pH ranging from 1 to 13. (\textbf{B}) Photo of the typical drop with a radius $R$ of 125 $\mu$m. The light spot at the center is caused by a light focusing on the front side of the microdrop illuminated from behind. The edge of the microdrop appears opaque due to total internal reflection. (\textbf{C}) Signal generated by the drop in FC. The charge of the drop is proportional to the area of the incoming or outgoing peak. (\textbf{D}) and (\textbf{E}) Charge of the drops from Kapton and Teflon capillaries as a function of applied voltage. The dots are experimental data, and the lines are linear regression curves. Colors correspond to different pH values. Each dot is an average of over 100 individual measurements. The error bars are about the size of the dots. $Q_0$ indicates the drop charge for a grounded electrode (Fig.~\ref{fig:2}). Note that charge is given in absolute values because the discussed below charging mechanism is independent of the drop volume.}
  \label{fig:1}
\end{figure}

We used a drop-on-demand generator (Fig.~\ref{fig:1}A) to produce uniform individual spherical microdrops of water with a radius $R$ (Fig.~\ref{fig:1}B). A reservoir of 10 milliliters was used as a  source of liquid with pH ranging from 1 (strong acid) to 13 (strong base), controlled by changing the concentration of HCl and NaOH in the wide range (see Methods). The drops were ejected from a cylindrical capillary by a piezo-electric actuator. All the microdrops had the same size within 10\% variation (Fig. S14 in the SI). We used two kinds of capillaries made of polyimide (Kapton) and polytetrafluoroethylene (Teflon). The charge $Q$ of a microdrop was measured by a Faraday cylinder (FC)\bibnote{Named by analogy with the Faraday cup: Brown, K. L.; Tautfest, G. W. Faraday-Cup Monitors for High-Energy Electron Beams. Rev. Sci. Instr. 1956, 27, 696–702}, consisting of the inner and the outer (grounded) shells (Fig.~\ref{fig:1}A). The voltage (Fig.~\ref{fig:1}C) induced by the charged drop moving along the axis of the FC inner shell was amplified and recorded by a digital oscilloscope. A voltage $U$ from 0 to $\pm$60 V was applied between the reservoir and the ground electrode by inserting a metal wire directly into the liquid. A video camera was used to image the microdrops and calculate their volume.

Figure~\ref{fig:1}, D and E, show an example of the dependence of the microdrop charge $Q$ on the applied voltage $U$ measured at different pH. The latter affects the slope of the lines, but not their linearity, which preserves in the whole $U$ range from 0 to $\pm$ 60 V. The slope of the lines gives the effective microdrop capacitance, $C_0$=$\Delta Q$/$\Delta V$ (see Fig. S18 in the SI). The capacitance equals approximately 0.015 and 0.025 pF for the drops from Kapton and Teflon capillaries, respectively. These values roughly coincide with the geometrical capacitance of the half sphere, $C_{sp}$=2$\pi \epsilon_0 \epsilon R$, where $\epsilon$ is the dielectric constant of the capillary, and $R$ is the capillary radius (see Methods). Thus, the charge is determined by the differential capacitance of the microdrop body at the point of separation. However, a detailed analysis shows that the charge $Q$ depends not only on the shape of the drop but also on the type of capillary material. Moreover, besides the direct charging of the drops with the external electrode potential, the microdrops are found to be charged even at $U$=0, and the charge does not depend on the drop volume. These facts together reveal some previously overlooked details of static electrification of water. 
 
\section*{Results and discussion}

\begin{figure}
    \centering
    \includegraphics[scale=0.45]{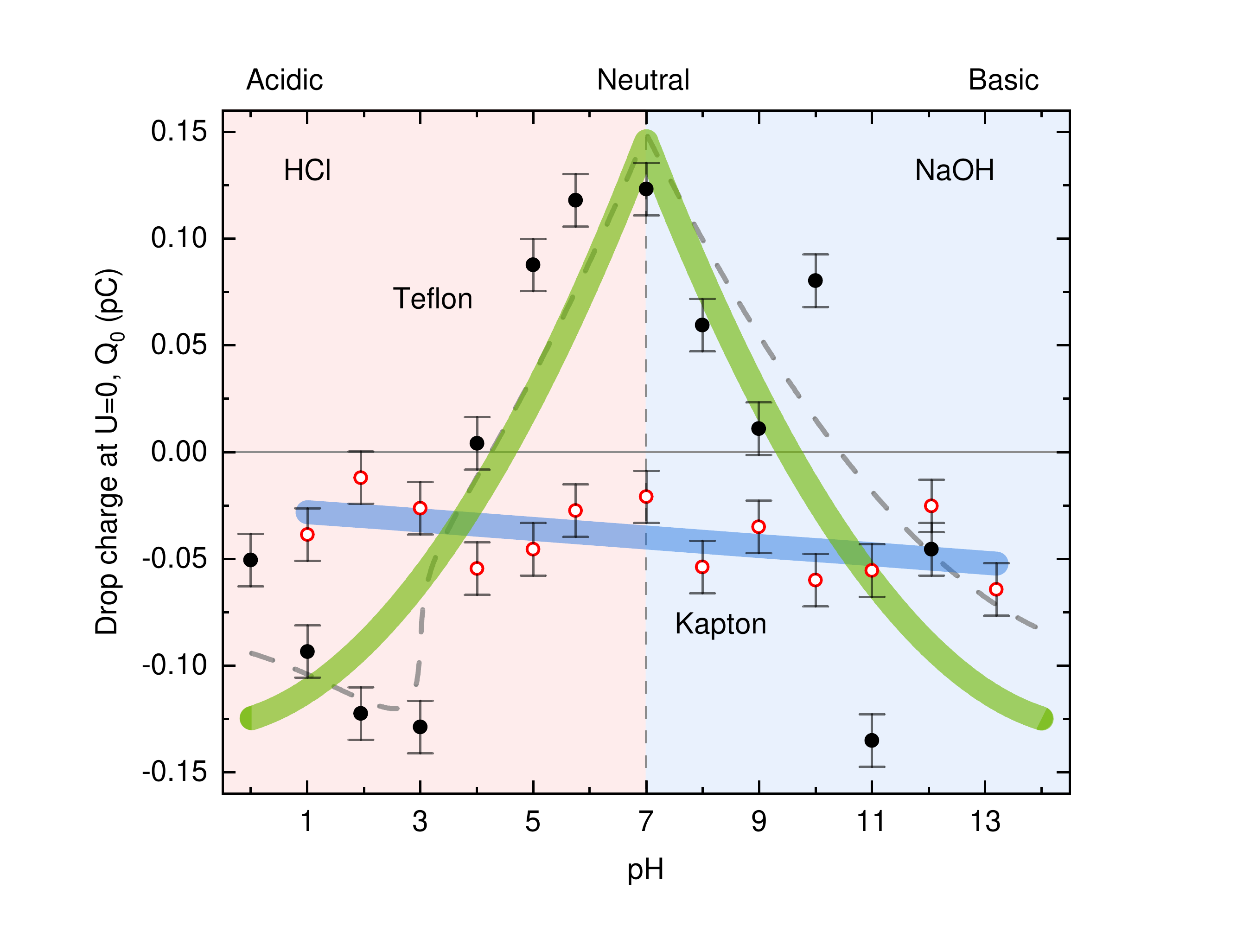}
    \caption{\textbf{Charge of water drops at different pH.} Open and closed dots are experimental data for Kapton and Teflon, respectively, and the lines are model functions: the linear approximation (blue), and the function according to Eq.\eqref{eq:Delta_phi_pH_U} (green). The dashed curve accounts for a linear pH dependence of the surface charge $\sigma$. Note that $Q_0 = Q_{tr} + Q_{off}$, where $Q_{off}$ = 0.12 pC is the offset charge due to the difference in the electrochemical potentials between liquid and electrode material, and $Q_{tr}$ is the transfer capacitance (see SI).}
\label{fig:2}
\end{figure}

Figure ~\ref{fig:2} shows nontrivial pH dependence of the microdrop charge $Q_0$ at $U$=0 different for different capillaries. The dependence of the charge on pH indicates that $Q_0$ is sensitive to the concentration of H$_3$O$^+$ and OH$^-$ ions in the solution. The capillary-type dependence indicates the importance of the distribution of the ions near the surface. Particularly, Kapton causes weak pH dependence (open dots and the blue line). All microdrops generated from a Kapton capillary are slightly negative. Teflon, on the other hand, yields drops whose charge has a remarkably strong bell-shaped pH dependence (closed dots and the green curve). The most striking feature of the microdrops generated from a Teflon capillary is the sign change: neutral-water drops from Teflon are positively charged ($Q_0^{max}$ = 0.15 pC) that corresponds to the previous findings~\cite{Nauruzbayeva2020}. But, at extreme pH values, the drops become negatively charged. The dependence is roughly symmetric around pH=7.

Both, the different shapes of the $Q_0$(pH) curves and the different amplitude of the charge variation for Kapton and Teflon capillaries indicate that there is an additional charging mechanism apart from the influence of an external field. This mechanism of zero-potential electrification is connected with the atomic-molecular structure of the liquid-solid interface, which is connected with the open water surface. Indeed, the electric potential at the metal electrode charges the whole body of water in which it is inserted. This effect is evident from the consistent relative shift of the curves shown in Fig.~\ref{fig:2} measured at different $U$ (see Fig. S2 in the SI). The parallel scaling of the curves shows that the shape of each $Q$(pH) dependence is irrelevant to the voltage $U$ (at least at relatively low electric field strength). It is determined by some other mechanism.

To shed light on the liquid-solid-air contact interactions, we performed independent contact-angle measurements. As a solid substrate, we used the same materials as that for capillaries. We found that the pH dependence of the contact angle of the drops on the Kapton surface is weak (Fig. S18B in the SI). The drops have nearly neutral behavior with almost 90-degree contact angle ($\theta \approx$ 85 $\pm$ 8$^{\circ}$). On the other hand, Teflon shows strong hydrophobic behavior ($\theta >$ 105$^{\circ}$) and displays a non-monotonic pH dependence of the contact angle (Fig. S18C in the SI). This dependence is symmetric around neutral pH and correlates with the pH dependence of the measured microdrop capacitance $C_0$ (Fig. S19 in the SI). Thus, the peculiarities of pH dependence of the charge of the drops from hydrophobic and neutral interfaces can be associated with the charge distribution near the three-phase liquid-solid-air interface. The symmetric form of both dependencies around neutral pH indicates that electrification strength depends on the solute concentration. 

\subsection*{Insights from MD simulations}

\begin{figure}
    \centering
    \includegraphics[scale=1.0]{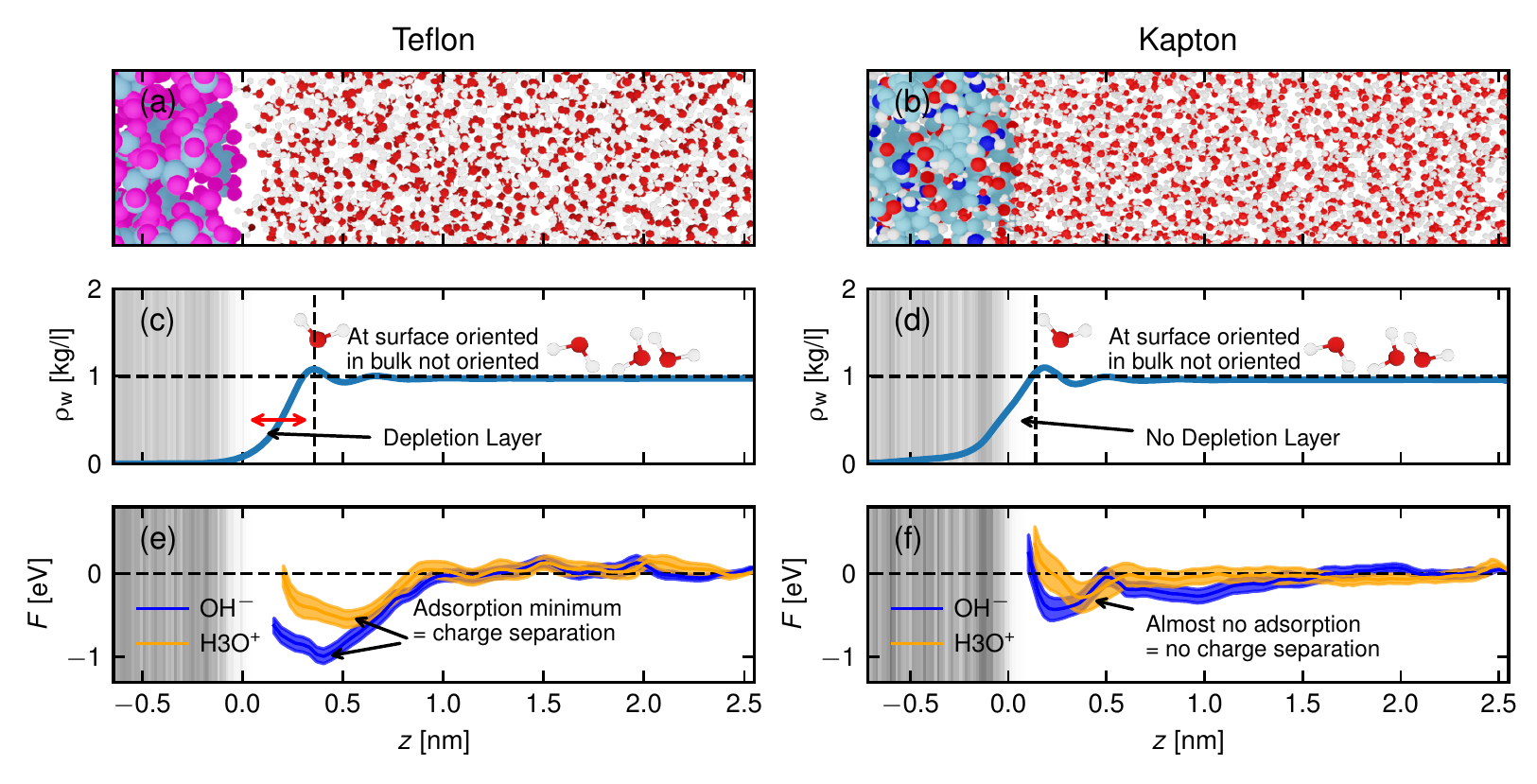}
     \caption{\textbf{Simulation of hydroxyl and hydronium adsorption at the capillary surfaces.} (\textbf{A}) and (\textbf{B}) Snapshots of the atomic distribution near Teflon-water (left) and Kapton-water (right) interfaces. Green, blue, red, white, and pink spheres are for carbon, nitrogen, oxygen, hydrogen, and fluorine atoms, respectively. (\textbf{C}) and (\textbf{D}) Density profiles for H$_2$O molecules. The solid phase is shown in grey. Vertical dashed lines are the Gibbs planes dividing solid and liquid phases (see SI). (\textbf{E}) and (\textbf{F}) Potential of mean force (PMF) profiles of hydronium (H$_3$O$^+$) and hydroxyl (OH$^-$) ions. Arrows show the minima of the distributions.}
\label{fig:3}
\end{figure}

To investigate the atomic-molecular structure of the liquid-capillary interfaces, we performed classical molecular dynamics (MD) simulations (see Methods and SI for the technical details). Figure~\ref{fig:3} summarizes the main results for the distribution of hydronium (H$_3$O$^+$) and hydroxyl (OH$^-$) ions near the capillary wall (results for Na$^+$ and Cl$^-$ are shown in Fig. S10 in the SI). The left and right columns correspond to Teflon and Kapton interfaces, respectively. The top panels (A and B) demonstrate snapshots of the atomistic distribution of pure water near the interfaces. The solid and liquid phases are shown on the left- and right-hand sides from $z=0$, respectively. The middle panels (C and D) show the density profiles for water molecules. The bottom panels (E and F) show the profiles of the potential of mean force (PMF), which correspond to the free energy surface of ions along the $z$ axis, for hydronium (H$_3$O$^+$) and hydroxyl (OH$^-$) ions.

Our MD simulations show differences in the distribution of ionic and molecular species near neutral Kapton and hydrophobic Teflon surfaces within the electrical double layer (z $\approx$ 0 - 2 nm). The H$_2$O molecules are randomly distributed in the diffuse layer (z $>$ 1 nm) but structured near the interface (surface polarization). The liquid phase is penetrating the Kapton wall, while the hydrophobic Teflon surface causes repulsion of the liquid by about 3 \AA (Fig.~\ref{fig:3}, C and D). This is obvious from the comparison of the location of Gibbs planes dividing solid and liquid phases (vertical dashed lines). Another difference is in the distribution of hydronium and hydroxyl ions (blue and yellow curves in panels E and F). Free-energy profiles of H$_3$O$^+$ and OH$^-$ ions show minima (adsorption of ionic species to the surface). The adsorption minima are deeper near Teflon assuming higher charge density and they are relatively shifted for positive and negative ions (charge separation). The ions of OH$^-$ are getting closer to the surface compared to ions of H$_3$O$^+$. Kapton, on the contrary, shows weak adsorption of ions with a negligible shift between positive and negative ionic densities. This behavior remains valid for Na$^+$ and Cl$^-$ ions (Fig. S10 in the SI). Thus, Kapton and Teflon differ in their ability to adsorb the charge and polarize liquid in their vicinity. Hydrophobic Teflon causes more profound charge separation than neutral Kapton. A higher affinity of negatively charged ions to the Teflon surface agrees with our experiments: the charge $Q$ on the microdrops at pH=7 is positive (Fig.~\ref{fig:2}), while the capillary is charged negatively.

\subsection*{The electrification mechanism}

\begin{figure}
    \centering
    \includegraphics[scale=1.05]{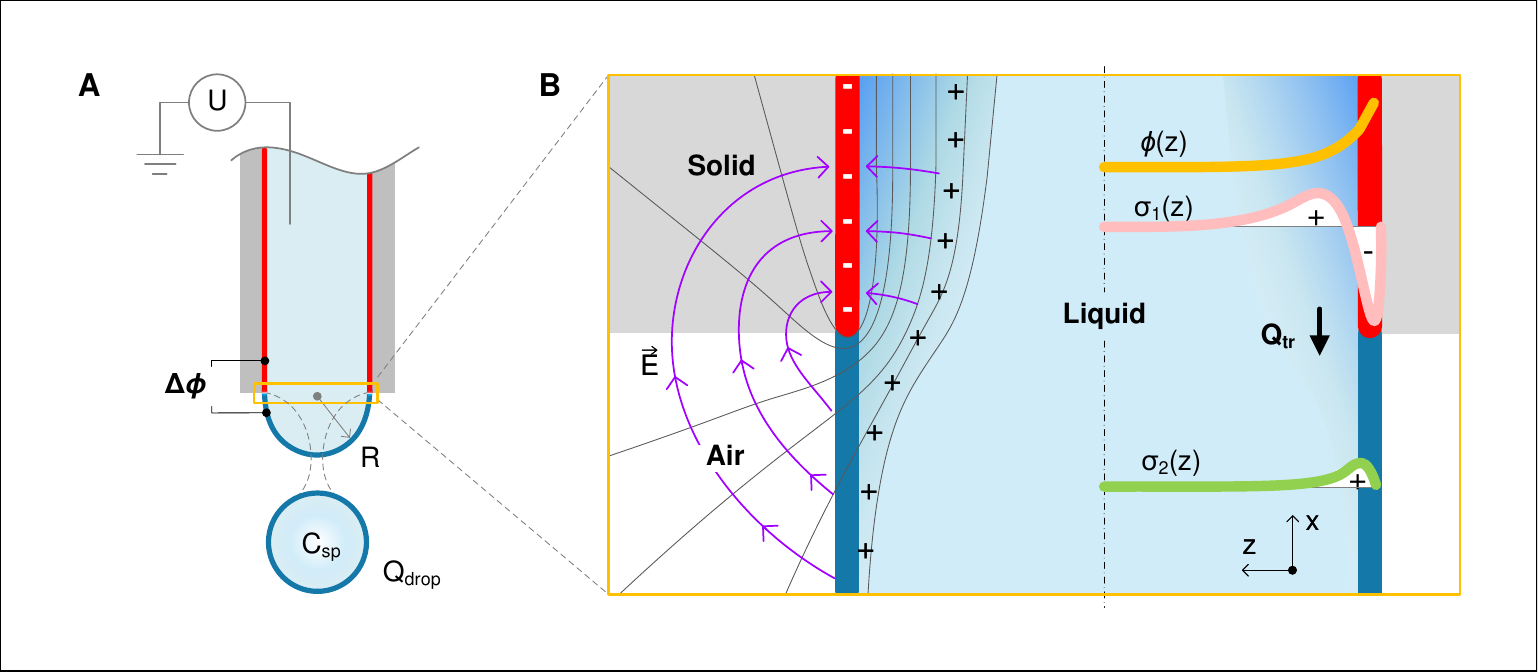}
    \caption{\textbf{Mechanism of water drop contact electrification.} (\textbf{A}) Capillary opening with emerging drop. The red line shows a negatively charged layer of ions adsorbed to the capillary surface. The blue line shows the water-air interface. $\Delta \phi$ is the contact potential difference caused by the different adsorption energy of ionic species of liquid to solid and air interfaces.(\textbf{B}) Schematic representation of the origin of the transferred charge. Electric field map near the edge of the capillary (left) and the radial distribution functions for the potential $\phi$ and the charge density $\sigma$ (right) along the z-axis at liquid-solid (pink) and liquid-air interface (green). Electric field lines and equipotentials are shown in magenta and black, respectively. In these boundary conditions, $\Delta \phi$ causes the charge transfer (black arrow), which settles on the drop surface.}
\label{fig:4}
\end{figure}

Experiments and simulations discussed above lead to the simple phenomenological model of the electrification mechanism. Figure~\ref{fig:4}A schematically shows the capillary and the microdrop before and after the detachment. The solid-liquid interface, which can adsorb charge, is shown in red. The liquid-air interface without adsorbed charge is shown in blue. We consider the electric potential difference $\Delta\phi$ between red and blue interfaces caused by the different adsorption/desorption energy of ions at the solid-liquid and liquid-air interfaces. The physical nature of $\Delta\phi$ is similar to that observed at the edges separating facets with different orientations of the terminating surfaces in crystalline solids or that at the junction of two metals with different electron work functions~\cite{Landau8}. The difference is only in the type of charge carriers, which for aqueous interfaces are ionic species. Note that $\Delta \phi$ differs from the streaming potential~\cite{Israelachvili2011}, which requires only one interface, while the contact potential difference discussed here requires contact of at least two interfaces. Note also that $\Delta \phi$ exists without any motion of the liquid. The timescale of the drop formation ($>$ 0.1 ms) is more than three orders of magnitude larger than the time needed for establishing the dynamic equilibrium for ionic species ($<$ 1 ns). Thus the considered electrification mechanism is essentially static. The potential difference $\Delta \phi$ is expected at any junction of two aqueous interfaces with a different adsorption energy of ions. 

To find the value of $\Delta \phi$, we consider the electrostatic distribution of the electric field lines and the charge carriers at the capillary edge (Fig.~\ref{fig:4}B). The negatively charged ions adsorbed on the solid surface are screened by the mobile counter-ions in the diffuse layer (pink curve). The electric field caused by this charge separation stays within the dipole layer characterized by the Debye screening length $\lambda$, which depends on the electrolyte concentration $n$, and the dielectric constant of the solution $\epsilon$ (see the SI). Adsorbed charge creates an electric potential drop between the surface and the bulk of the liquid (yellow curve). Since the electrolyte-air interface has a negligible adsorbed charge compared to that at the solid surface, the contact potential difference $\Delta \phi$ along the $x$ axis equals the potential drop across the dipole layer perpendicular to the solid-liquid boundary along the $z$ axis. The potential difference between red and blue interfaces stimulates charge transfer across their contact, which eventually means the charging of the drop. Most of the charge settles at the distance larger than $\lambda$ from the capillary-drop contact line.

The amount of the transferred charge $Q_{tr}$ between the capillary and the microdrop is equal to $C_{tr}\Delta \phi$, where $C_{tr}$ is the transfer capacitance of the charge depletion zone (see Sec. 1.1 in the SI). The capacitance $C_{tr}$ is independent of the Debye screening length, and the pH dependence of $Q_{tr}$ is mainly due to the change of $\Delta \phi$. The latter may depend on pH for two reasons. The first is the change in the density of the adsorbed charge $\sigma$, which is expected to be asymmetric around pH = 7~\cite{Israelachvili2011}. The second is the change in the screening length $\lambda$, which is symmetric around pH = 7. Note that $\lambda$ varies within three orders of magnitude between pH=1 and pH=13 with the maximum at pH=7. The nearly-symmetric experimental dependence of $Q_0$ on pH (Fig.~\ref{fig:2}) suggests that it is mainly determined by the change of $\lambda$ rather than $\sigma$.  

The Poisson-Boltzmann formalism (see Sec. 1.2 in the SI) for the geometry shown in Fig.~\ref{fig:4} gives the contact potential difference between solid-liquid and liquid-air interfaces:
\begin{align}
    \label{eq:Delta_phi_pH_U}
    \Delta \phi(pH) = & \phi_0 \sinh^{-1} \left( \frac{\lambda(pH) \sigma}{\phi_0\epsilon_0} \right), 
\end{align}
where $\phi_0 = 2 k_BT/q$ is the saturation value at high $\sigma$, and $\lambda (pH)$ is the pH-dependent Debye length, given by $\lambda_0 10^{-|pH-7|/2}$. The fit of Eq.~\eqref{eq:Delta_phi_pH_U} to the experimental data is shown in Fig.~\ref{fig:2} (green line). The best fit parameter is $\lambda_0 \cdot \sigma$ = 1.3$\cdot$10$^{-8}$ C/m. Note that assuming the simplest linear pH dependence of the adsorbed charge density $\sigma = \sigma_0 (pH-pK_a)$, where $pK_a$ = 3~\cite{Barisic2019}, we get further improvement of the experimental data fitting (dashed gray line in Fig.~\ref{fig:2}). The study of the pH dependence of $\sigma$, however, goes beyond the current research.

In conclusion of this section, we would like to point out an important feature of the contact voltage induced by the charge separation at the junction of two aqueous interfaces. The dependence given by Eq.~\eqref{eq:Delta_phi_pH_U}  shows saturation of the contact potential difference $\Delta\phi$ at large $\sigma$. This means that for surfaces with strong charge separation, such as Teflon, the contact potential $\Delta\phi$ becomes insensitive to changes in pH and other system parameters. In this case, the contact potential reaches a universal saturation value determined by the temperature and the charge $q$ only:
\begin{align}
\label{eq:contact_potential_universal} \phi_0 = & 
\frac{2 k_B T }{q} = \frac{2\cdot 1.38 \cdot 10^{-23} [m^2kg \cdot s^{-2} K^{-1}]\cdot 300 [K] }{1.6 \cdot 10^{-19} [C]}   \approx 0.052~V.
\end{align}
This value represents a universal potential drop, which paves the way for a mechanism of voltage stabilization, $\Delta\phi \rightarrow \phi_0$, induced by interface charge adsorption and separation. Since it arises in aqueous solutions it may play an important role in biological systems. To our knowledge, the existence of this mechanism of voltage stabilization in biological systems, and its possible relevance to biological function have not been explored.

\section*{Conclusions}

We have revealed a previously overlooked static mechanism of electrification of aqueous interfaces by the precise generation of liquid microdroplets with different pH values. The mechanism consists of the formation of contact potential difference $\Delta \phi$ between the liquid-solid and liquid-air interfaces with different amounts of adsorbed charge. Our phenomenological model and theory quantitatively describe the experimental data and assume the universality of the phenomenon that makes it relevant to biological sciences, nanofluidics, and electrochemical energy storage. The immediate application of the observed phenomenon is an electrodeless pH determination. In general, our research answers the long-lasting question about the charge of the open-water surface, whose value is shown to depend on the properties of the contact between the water-solid and water-air interface and can be both positive or negative. The best recipe for having drops without charge is to use neutral materials with a contact angle close to 90$^{\circ}$ and equal energy of surface adsorption for positive and negative ions. To maximize the drop charge (e.g., for triboelectric energy harvesting) one needs to use highly hydrophobic (e.g., Teflon), or highly hydrophilic (e.g., cellulose) materials and pH-neutral water. Further studies on the contact potential difference at the junction of various aqueous interfaces may provide a better understanding of biological potentials and pave the way for the improvement of electrochemical power systems, sensors, ionotronics, and energy harvesting devices.

\section{Methods}

\subsection{Microdrops generation and sample preparation}

We used a commercial drop-on-demand (Biofluidics) generator of microdrops. We used pure 18-MOhm water as a buffer, and solutions of HCl (acid), and NaOH (base) with concentrations from 10$^{-7}$ M to 10$^{-1}$ M. Solutions were prepared by mixing the corresponding amount of concentrated liquid with buffer just before the experiments. pH of the mixtures was controlled by a glass-electrode pH meter for each sample after the preparation. The external electric field was created by applying voltage $U$ between the nozzle and the ground. We used platinum and copper electrodes, which were directly placed in the solution (Fig.~\ref{fig:1}A). No significant effect of metal type on the microdrop charge was found (Fig. S13 in the SI). The microdrop shape and size (Fig. S11 in the SI) were controlled by a video camera synchronized with a piezo injector. We used two commercial (Biofluidics) capillaries made of Polytetrafluoroethylene (Teflon) and Polyimide (Kapton) of the diameters of 125 $\mu$m and 200 $\mu$m, respectively (Fig. S12 in the SI). The contact-angle measurements (Fig. S18 in the SI) were made on the same materials as that of capillaries.

\subsection{Droplet charge measurements}

The charge $Q$ of microdrops was measured at different voltages $U_{\pm}$ ranging from 0 to $\pm$ 60 V. The voltage was applied between the injector capillary and the ground electrode (Fig.~\ref{fig:1}A). The charge $Q$ was registered in the Faraday cylinder by using a commercial current amplifier (FEMTO) and the oscilloscope (Tektronix). The measured $Q$ values of individual microdrops were averaged over 100 measurements for each pH value at the voltage $U_{\pm}$. As a result, the accuracy of the $Q$ measurements was better than 0.01 nC. The voltage $U_{\pm}$ was always well below the Rayleigh stability limit~\cite{Duft2003} (see Fig. S16 in the SI). No influence of the electric field on the microdrop stability was observed (Fig. S15 in the SI). The charge $Q$ of microdrops was provided in absolute charge units (see, e.g., Figs.~\ref{fig:1}, and ~\ref{fig:2}). We did not normalize the charge by the drop volume $V$ because the discussed charging mechanism does not assume its dependence on $V$ or the drop surface area $S$ (see Sec. 1 in the SI). On the contrary, it depends only on pH and the length of the capillary perimeter $\textit{l}=2\pi R$ (the contact line). As the diameter of the Kapton capillary was 1.6 larger, we divided $Q$ from this capillary by a factor of 1.6 in Fig.~\ref{fig:2} to allow for the parallel comparison of the data. 

\subsection{Molecular dynamics simulations}

Classical molecular dynamics (MD) simulations were performed using the GROMACS~\cite{Abraham2015} software of version 2021.5. For H$_2$O molecules we used the SPC/E model. For hydronium and hydroxide ions, we used the thermodynamically consistent force field described elsewhere\cite{Bonthuis2016}. Our unit simulation box included a Kapton/Teflon slit pore of 4 nm $\times$ 4 nm $\times$ 19 nm, which was filled with water. This geometry enabled us to model both the distribution of species close to the surface and the water structure in the capillary. To generate the Kapton/Teflon walls, we followed a previously described procedure\cite{Wloch2018}, which was shown to reproduce the contact angles of water drops. Using the Automated Topology Builder\cite{Stroet2018}, we first generated an atomistic geometry of the molecule (Fig.S1 in the SI). Then, we constructed the preliminary wall surface by compressing 100 molecules into a single slab of 4 nm$^2$ in lateral dimensions. To verify the result we compared the calculated surface roughness with the experimental values~\cite{Wloch2018}, which were found to be in good agreement. Finally, a water slab of 15 nm was placed between the two pre-generated wall structures.

To investigate the difference between Kapton and Teflon interfaces at different pH values, we performed free energy calculations. The axis perpendicular to the capillary surface was chosen as a reaction coordinate and the potential of the mean force profile (PMF) was calculated. To improve sampling, we utilized the umbrella sampling technique\cite{Kastner2011}, which generates the free energy profile for a small window around the particular reaction coordinate. Multiple windows were combined using the method of weighted histogram analysis (WHAM) \cite{Kumar1992, Souaille2001, Hub2010}, which optimizes the weight of each window in the combined PMF to minimize the variance of the estimated profile. We used the bootstrap method (see SI) to estimate the standard error of the mean of the PMF profile and plot the error bars. Further technical details, simulation input scripts, and additional simulation results are available in the SI.

\begin{acknowledgement}

We thank Ilya Kopanichuk, Pavel Kapralov, Henni Ouerdane, and Alexander Ryzhov for fruitful discussions; Philip Loche for the help with setting up simulation systems and the determination of free energy profiles; Nathan Ronceray and Charles Asbury for the careful reading of the manuscript.

The work of A. S. and P. S. is funded by Deutsche Forschungsgemeinschaft (DFG, German Research Foundation) under Germany's Excellence Strategy - EXC 2075–390740016. They acknowledge the support of the Stuttgart Center for Simulation Science (SimTech). The work of A. A. is supported, in part, by the US National Science Foundation through the MRSEC grant  DMR-1719797, the Thouless Institute for Quantum Matter, and the College of Arts and Sciences at the University of Washington. V. A. acknowledges EPFL support (base funding), and A. R. is funded by the Swiss National Science Foundation (SNSF) through 200021-192037.

\end{acknowledgement}

\begin{suppinfo}
The supplementary information is available online, including the theory of the charge transfer between two dissimilar aqueous interfaces induced by the contact potential difference, details of molecular dynamics simulations, and additional graphs, photos, and diagrams, including Figures S1 to S18 and Table S1.
\end{suppinfo}

\bibliography{references}
\end{document}

\textbf{This PDF file contains:}

1. Charge transfer between two dissimilar aqueous interfaces induced by the contact potential difference

2. Molecular dynamics simulations

3. Additional graphs, photos, and diagrams

•	Figures S1 – S18

•	Tables S1

\newpage

\section{1. Charge transfer between two dissimilar aqueous interfaces induced by the contact potential difference}

We consider electrolyte solutions in which the solute is dissociated into ions. The solution is placed into a cylindrical capillary with an open end (see Fig. 4A in the main text). There are two aqueous interfaces in the system: the capillary-electrolyte and the electrolyte-air (the latter can be any other dielectric). When two different substances come into contact, charge separation generally occurs and electric potential difference develops across the interface. In the case of an electrolyte, charge separation is driven by the adsorption of ions at the interface (in the Stern layer) and the screening of adsorbed charge by the counterions in the diffuse layer (see Fig. 4B in the main text). The properties of the dipole layer depend on the material in contact with water. Since water-capillary and water-air interfaces are different, a potential difference  $\Delta \phi$ develops at the contact line between them. This potential drop $\Delta \phi$ leads to charge transfer $Q_{tr}$ between the two surfaces, resulting in the charging of the initially charge-neutral interfaces.
Here, we present a theory that quantitatively describes the experimental data obtained in this article.

\begin{figure}[h]
    \centering
    \includegraphics[width=0.90\textwidth]{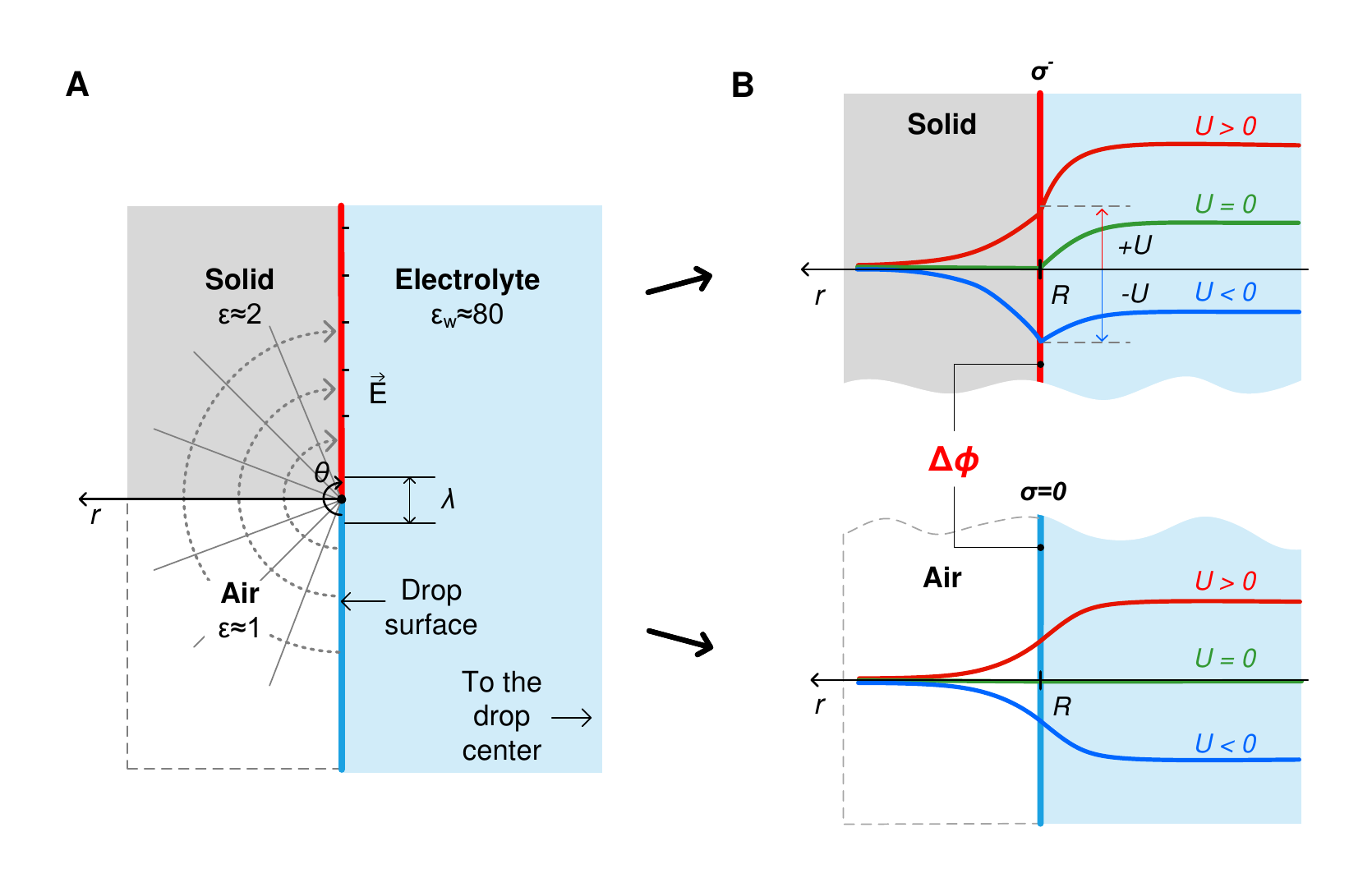}
    \caption{\textbf{Electrostatics of triple-phase interface.} (\textbf{A}) The water-air-capillary point of contact. The red and blue lines are interfaces with and without adsorbed charge (open-water surface), respectively. Dashed lines represent the distribution of the electric field $E$. $\theta$ and $r$ are polar coordinates. $\lambda$ is the Debye screening length. (\textbf{B}) Potential distribution near solid-electrolyte (top) and air-electrolyte (bottom) surface at a distance much larger than $\lambda$. $U$ is the electric potential due to applied voltage (positive or negative). $\sigma$ is the adsorbed charge surface density. $\Delta\Phi$ is the contact potential difference.}
    \label{fig:T1}
\end{figure}

\subsection{1.1 Transfer capacitance}
\label{sec:transfer_capacitance}

The contact potential difference results in charge transfer across the contact line. 
The transferred charge $Q_{tr}$ may be expressed in terms of the transfer capacitance as:
\begin{align}
    \label{eq:Q_tr_C_tr}
    Q_{tr} = & \, C_{tr} \, \Delta \phi.
\end{align}
The value of the transfer capacitance $C_{tr}$ depends on the geometry of the capillary, the capillary dielectric constant $\epsilon$, and the shape of the drop at the moments of separation. Its determination requires solving the electrostatic problem outside the water surface with the potentials at the liquid-solid and liquid-air interfaces set to $\Delta \phi$ and zero, respectively. In this setup, the surface density of the transferred charge is inversely proportional to the distance to the contact line, and the transfer capacitance is dominated by the spatial region near the contact line. In this region, the system can be modeled by wedges of water, air, and the dielectric joined at the contact line.

Figure~S\ref{fig:T1}A shows a point at the contact line between liquid-air and liquid-solid interfaces. The air and the solid part of the system are represented by wedges with opening angles $\theta$ and $\theta_\epsilon$, respectively, where $\epsilon$ is the dielectric constant of the capillary. In the shown right-angle geometry, $\theta = \theta_\epsilon = \pi/2$. The contact potential difference $\Delta \phi$ between the electrolyte-solid (red) and the electrolyte-air (blue) interfaces creates an electric field $E$, whose field\sout{s} lines go in the azimuth direction and equipotential surfaces emanate from the contact line at different azimuth angles. The magnitude of the electric field $E$, and the density of the charge transferred from red to blue surfaces are inversely proportional to the distance $r$ from the contact point.

Integrating the surface charge density between the inner radius,  $R_{in}$, and  outer radius, $R_{out}$, of the wedge,   we get the transfer capacitance per unit length of the contact line to be $\frac{4\pi  \epsilon_0}{\theta +\theta_\epsilon/\epsilon} \ln \frac{R_{out}}{R_{in}}$. For our systems $R_{out}$ is about the capillary radius, whereas $R_{in}$ is about the Debye screening length $\lambda =\sqrt{\frac{k_BT \epsilon_w \epsilon_0}{2 n q^2}}$, where $n$ is the electrolyte concentration, and $\epsilon_w$ is the dielectric constant of water. $\lambda$ is small, thus, the logarithmic factor $\ln \frac{R_{out}}{R_{in}}$ is large. Most of the transferred charge resides at distances $r \gg \lambda$ from the contact line. At such distances, the density of the transferred charge is small compared to that on the solid-liquid interface. Therefore, the contact potential difference $\Delta \phi$ is practically unaffected by the transferred charge, which justifies the validity of Eq.~\eqref{eq:Q_tr_C_tr}. 
Taking into account that the length $l$ of the contact line is equal to $2\pi R$, we have the transfer capacitance:
\begin{align}
\label{eq:C_tr_log}
    C_{tr}  = &  \frac{8 \pi^2 \epsilon_0 R}{\theta + \theta_\epsilon/\epsilon } \ln \frac{R}{\lambda}.  
\end{align}
Taking $\theta_{air} = \theta_\epsilon = \pi /2$ we get the following estimate for the transfer capacitance:
\begin{align}
\label{eq:C_tr_estimate}
    C_{tr}  \sim  &  \frac{16 \pi \epsilon_0 \epsilon R}{1 +\epsilon} \ln \frac{R}{d}.  
\end{align}
A more realistic geometry of the drop at the moment of the separation is shown in Fig. 4A of the main text (see dashed line). In this geometry, only the charge below the pinch-off line is carried away by the drop. Therefore, the logarithmic factor in the above expression can be roughly replaced by unity. In this case, $C_{tr}$ is about 0.06 pC for $R$ = 125 $\mu$m and $\epsilon \approx$ 1.

\subsection{1.2 Dependence of the contact potential on pH}
\label{sec:pH_dependence}

\begin{figure}
    \centering
    \includegraphics[scale=0.40]{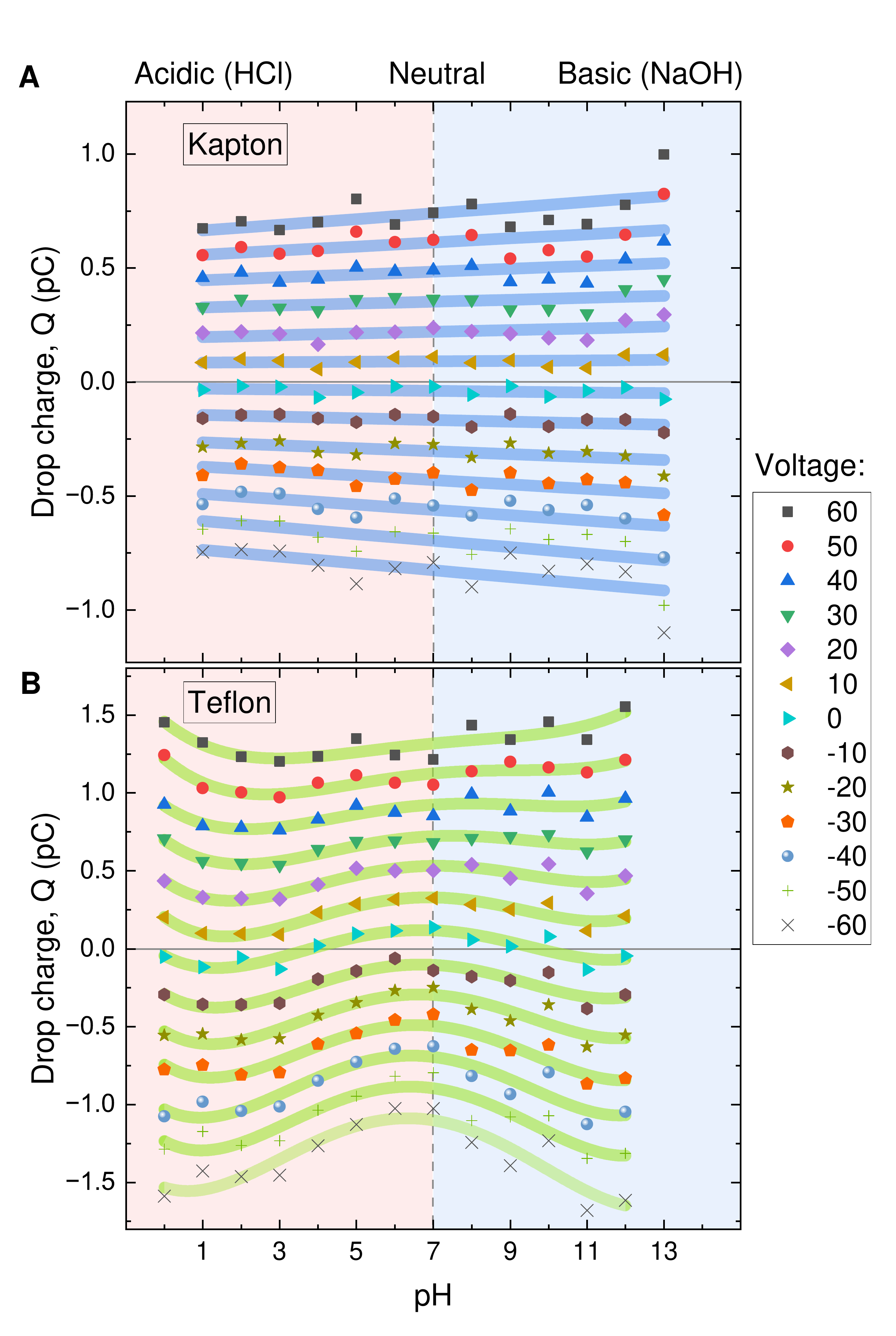}
    \caption{\textbf{Dependence of the drop charge on pH for different voltages.} (\textbf{A}) Kapton capillary. (\textbf{B}) Teflon capillary. The dots are experimental data from Fig. 1, D and E (main text), and the lines are linear (top) and polynomial (bottom) approximations, which are guides for the eye.}
\label{fig:SI_10}
\end{figure}

Equation \eqref{eq:Q_tr_C_tr} expresses the transferred charge $Q_{tr}$ between the liquid-solid and liquid-air surfaces in terms of the geometric transfer capacitance of the drop $C_{tr}$ and the contact potential $\Delta \phi$. There are two reasons why $Q_{tr}$ can depend on pH. The first is the change in the density of the adsorbed charge $\sigma$, which changes asymmetrically around pH = 7~\cite{Barisic2019}. The second is the change of the screening length $\lambda$, which is, in contrast, symmetric around pH = 7. Experimental data show that when the electrode is grounded ($U$ = 0), $Q_0$ exhibits a bell-shaped dependence on the pH of the solution, which is nearly symmetric around the neutral water with pH = 7 (Fig. 2 in the main text). Upon application of a negative voltage $U$ to the electrode this dependence is enhanced, while the application of positive $U$ diminished the amplitude of the bell-shaped pH dependence (Fig.~S\ref{fig:SI_10}). This behavior suggests that the pH dependence is mainly due to the change of $\lambda$ rather than the change of $\sigma$. In this section, we provide a simple theory, which gives a quantitative dependence of the contact potential $\Delta \phi$ on pH and $U$.

The driving mechanism of the contact potential difference $\Delta\phi$ is the charge double layer that is formed at the liquid-capillary surface. As discussed in Sec. 1.1 above, we neglect the influence of the transferred charge on $\Delta \phi$. The latter is determined by evaluating the electric potential drop across the charge double layer for infinite water-air and water-solid interfaces, and taking the difference between them (Fig.~S\ref{fig:T1}B).  To do this, we set the electric potential $\phi_{\infty}$ in the bulk of the liquid to zero and evaluate the potential $\phi(R)$ at the surface of the drop. The latter depends on the surface density of the adsorbed charge and the electrode voltage $U$. To describe the dependence on $U$, we consider the simplest geometry of a spherical drop of radius $R$. Inside the diffuse layer in the drop the dependence of the electric potential on the radial coordinate $r < R$ obeys the Poisson-Boltzmann (PB) equation. In  the case of electrolyte consisting of singly charged cations and anions, $q= |e|$, the Langmuir solution~\cite{Langmuir1938} of the PB equation is:
\begin{align}
    \label{eq:phi_inside}
    \tanh \frac{q \phi (r)}{4 k_B T} = & \tanh \frac{q \phi(R)}{4 k_B T} \exp \frac{r -R}{\lambda}.
\end{align}

Outside the drop, the electric potential is given by Coulomb's law:
\begin{align}
    \label{eq:phi_outside}
    \phi (r) = \phi(R) + U \left(\frac{R}{r} -1 \right),
\end{align}
where $U$ is the potential difference between the drop surface and the ground.

The discontinuity of the electric field at the interface of the drop (Fig.~S\ref{fig:T1}B) is related to the adsorbed surface density $\sigma$ in the Stern layer by Gauss' law. Using Eqs.~\eqref{eq:phi_inside} and \eqref{eq:phi_outside} we get:
\begin{align}
    \label{eq:phi_Gauss_law} 
    \frac{\sigma}{\epsilon_0} = & \frac{U}{R} + \frac{2 k_B T}{q\lambda} \sinh \frac{q \phi (R) }{2 k_B T},
\end{align}
which yields the potential at the surface of the drop:
\begin{align}   
    \phi(R) = &  \frac{2 k_B T}{q} \sinh^{-1} \left( \frac{q \lambda }{2 k_B T } \left[\frac{\sigma}{\epsilon_0} - \frac{U}{R} \right] \right). 
\end{align}

We note that a nonzero density of adsorbed charges is generically expected at any surface between the electrolyte and a different substance. Broadly speaking one can distinguish between surfaces with strong and weak charge separation. For strongly adsorbing surfaces the argument of the hyperbolic arcsine is large, $ \frac{ \lambda q \sigma}{2 \epsilon_0 k_B T } \gg 1$. In this case, the logarithmic dependence of the surface potential on the density $\sigma$ of the adsorbed charge is weak, and the surface potential  reaches a universal limiting value $\sim 2 k_B T/|q|$, controlled only by the temperature. In the opposite case of weak adsorption of ions on the surface, the surface potential $\phi(R)$ exhibits a linear dependence on the surface density of charge $\sigma$.

Our data indicate that $\sigma$ at the free water surface is negligibly small compared to that at the capillary wall. Therefore, the contact potential difference is given by:
\begin{align}
    \label{eq:Delta_phi_U}
    \Delta \phi(\lambda, U) = &  \frac{2 k_B T}{q} \left\{ \sinh^{-1} \left(  \frac{q \lambda }{2 k_B T } \left[\frac{\sigma}{\epsilon_0} - \frac{U}{R} \right] \right)  + \sinh^{-1} \left( \frac{q \lambda  }{2 k_B T} \frac{U}{R } \right)  \right\},
\end{align}
where both, the screening length $\lambda$, and the surface charge density $\sigma$ can depend on pH. We note that the PB equation assumes collective screening, and is expected to apply sufficiently far away from neutral pH, where the concentration of ions  $n_+ = n_- = n$ is sufficiently high.   In this case,  the screening length is inversely proportional to the square root of the concentration, for a solution of monovalent ions is given by $\lambda =\sqrt{\frac{k_BT \epsilon_w \epsilon_0}{2 n q^2}}$, where $n$ is the electrolyte concentration, and $\epsilon_w$ is the dielectric constant of water. Thus, not too close to pH$=7$ the pH dependence of the screening length is given by:
\begin{align}
    \label{eq:lambda_pH}
    \lambda = & \lambda_0 10^{-\frac{|pH - 7|}{2}},
\end{align}
where $\lambda_0 \approx $  1 $\mu$m. Substituting Eq.~\eqref{eq:lambda_pH} into Eq.~\eqref{eq:Delta_phi_U} we obtain the following dependence of the contact potential on pH and the electrode potential $U$:
\begin{align}
    \label{eq:Delta_phi_pH_U}
    \Delta \phi(pH, U) = & \phi_0 \left\{ \sinh^{-1} \left( \phi_0^{-1}\lambda_0  10^{-\frac{|pH-7|}{2}} \left[\frac{\sigma}{\epsilon_0} - \frac{U}{R} \right] \right)  + \sinh^{-1} \left( \phi_0^{-1}\lambda_0 10^{-\frac{|pH-7|}{2}} \frac{U}{R} \right)  \right\}. 
\end{align}

As mentioned above, the description of the pH dependence of the contact potential in Eq.~\eqref{eq:Delta_phi_pH_U} applies for sufficiently large ion concentrations where screening is collective, and breaks down  only in the limiting case in the vicinity of pH = 7. Note that since the screening radius is an even function of pH - 7, substitution of Eq.~\eqref{eq:Delta_phi_pH_U} into Eq.~\eqref{eq:Q_tr_C_tr} produces a symmetric dependence of the transferred charge on $|$pH $-7|$. This gives a good description of the experimental data. We believe that the slight asymmetry of the experimental bell-shaped curve shown in Fig. 2 (main text) can be explained by further accounting for the pH dependence of the surface charge $\sigma$. The quantitative treatment of this effect is beyond the scope of the present study.

The fit of Eq.~\eqref{eq:Q_tr_C_tr} with $\Delta\phi$ determined by Eq.~\eqref{eq:Delta_phi_pH_U} to the experimental data is shown in Fig. 2 of the main text (green line). The only best-fit parameter is $\lambda_0 \cdot \sigma$ = 1.3$\cdot$10$^{-8}$ C/m. Assuming $\lambda_0$ = 1 $\mu$m, we get $\sigma$ = 0.01 C/m$^2$, which is close to the experimental value of the charge density at the inner layer of the Teflon-water interface~\cite{Preocanin2012}. Equation \eqref{eq:Delta_phi_pH_U} also quantitatively described the dependencies on $U$ shown in Fig.~S\ref{fig:SI_10}. Though, the deep analysis of the external field effect on the water drop charge goes beyond the current study.

\section{2. Molecular dynamics simulations}

All simulations were performed using the GROMACS software package\cite{Abraham2015}, version 2021.5. All the input scripts for the simulations are accessible on DARUS\cite{data_staerk2022a}.

\subsection{2.1 System setup and structure generation}

\begin{figure}[h]
    \centering
    \includegraphics[width=.35\textwidth]{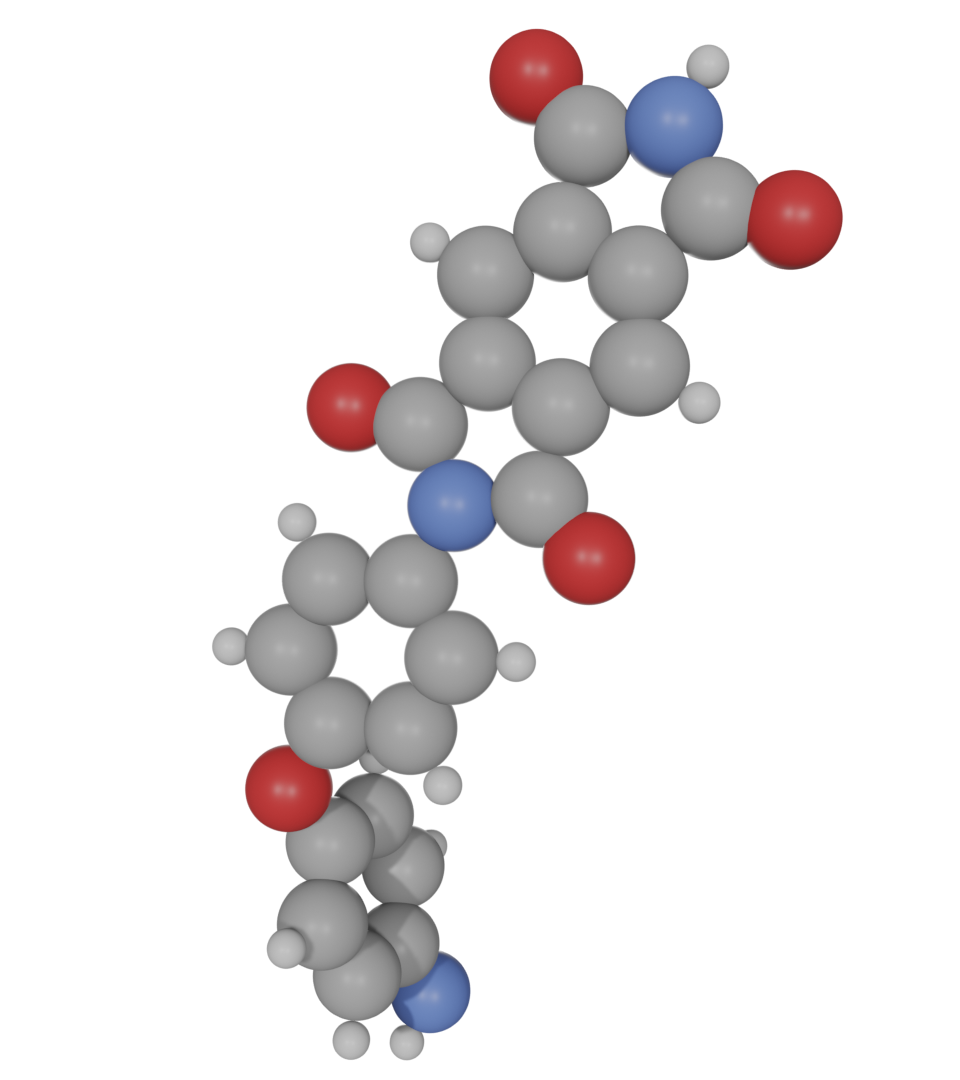}
    \includegraphics[width=.35\textwidth]{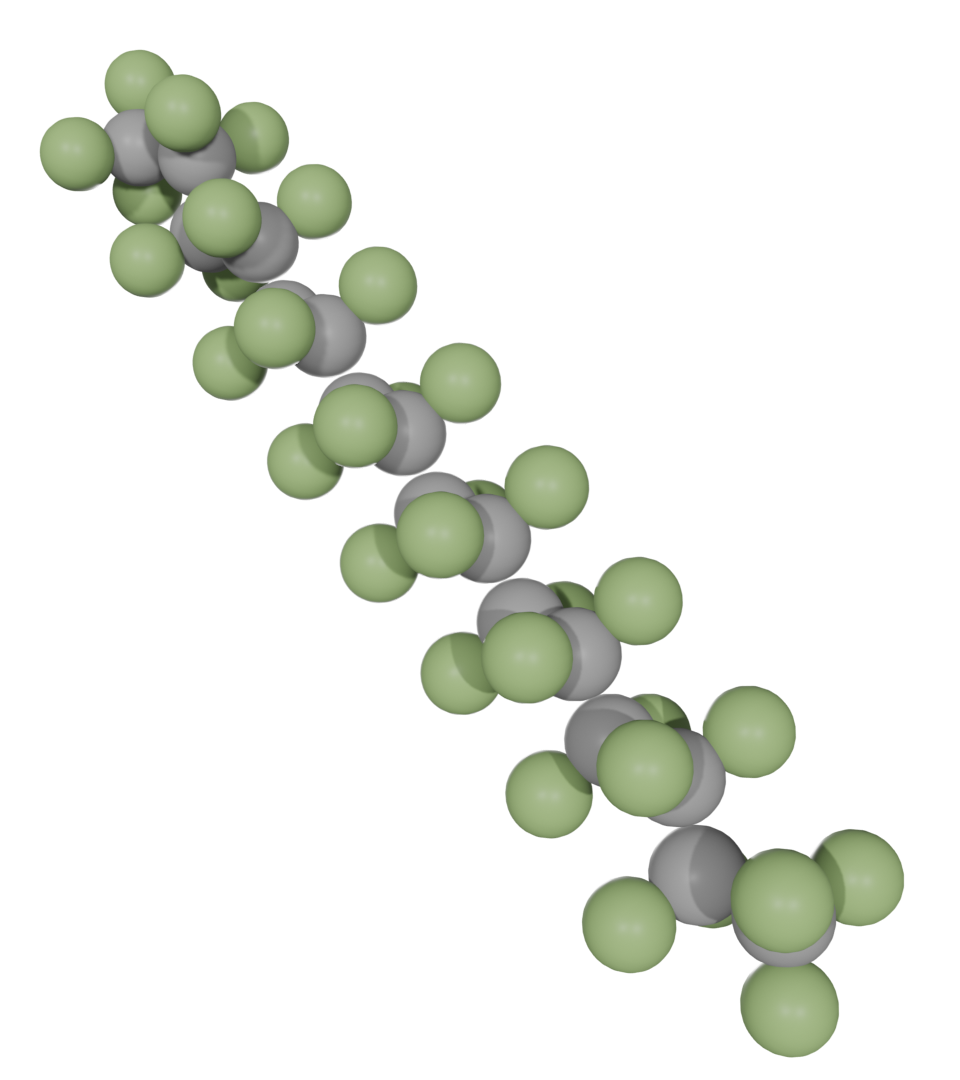}
    \caption{Snapshot of the individual molecule's structures that were used in the simulations: the Kapton molecule (left) and the PTFE (Teflon) molecule (right). Grey, green, blue, and red colors are for carbon, fluorine, nitrogen, and oxygen atoms, respectively.}
    \label{fig:MD1}
\end{figure}

To generate the structure of Kapton and Teflon (Polytetrafluoroethylene, PTFE), we utilized the automated topology builder to get optimized geometries of Kapton and Teflon molecules and obtain the information on the partial charge (Fig. S\ref{fig:MD1}). The structures used for the simulations in this paper can be found under \href{https://atb.uq.edu.au/molecule.py?molid=900833}{https://atb.uq.edu.au/molecule.py?molid=900833} (Kapton) and \href{https://atb.uq.edu.au/molecule.py?molid=352600}{https://atb.uq.edu.au/molecule.py?molid=352600} (Teflon). To generate surfaces comparable to the capillaries used in the experiments, we utilized a strategy detailed by Włoch et al.\cite{Wloch2018}. In short, high pressure is applied along the z-axis under high temperature, effectively melting the polymers and compressing the melt into a slab. For both Kapton and Teflon, we used 100 molecules inside the $x\times y$ = 4 nm $\times$ 4 nm simulation domain with periodic boundary conditions in the $x, y$ directions. Simulations for the melting and compacting step were performed using GROMACS with an initial temperature of 1000 K and a pressure of 1000 bar (semi-isotropic, Berendsen barostat) applied in the $z$ direction coupled to 9-3 Lennard-Jones walls. In four subsequent steps, the structure was quenched first at 800 K followed by 600 K, 500 K, and finally 300 K, i.e. the target temperature for the simulations.

\begin{figure}[h]
    \centering
    \includegraphics[width=0.49\textwidth]{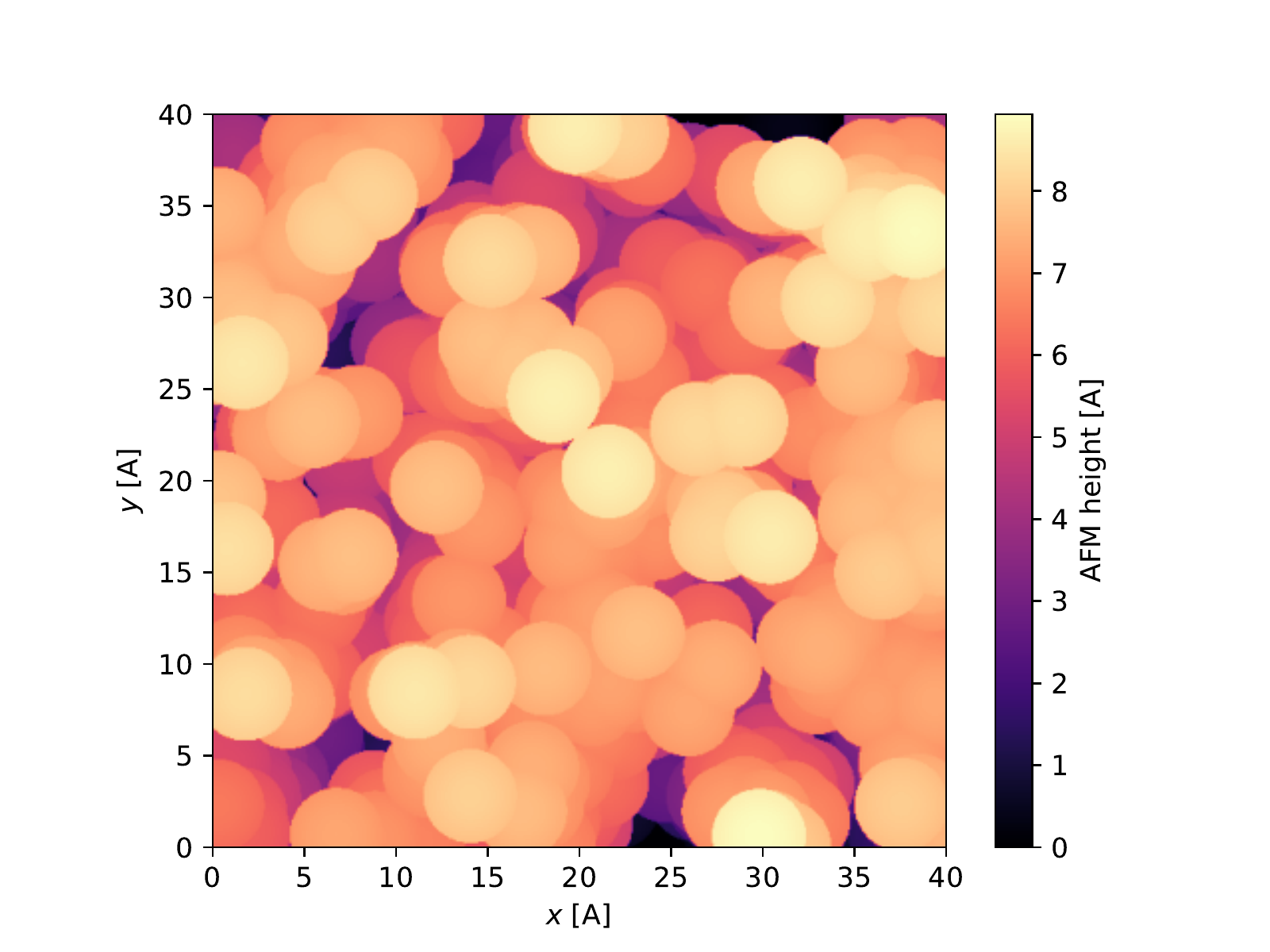} % Kapton is nr.2
    \includegraphics[width=0.49\textwidth]{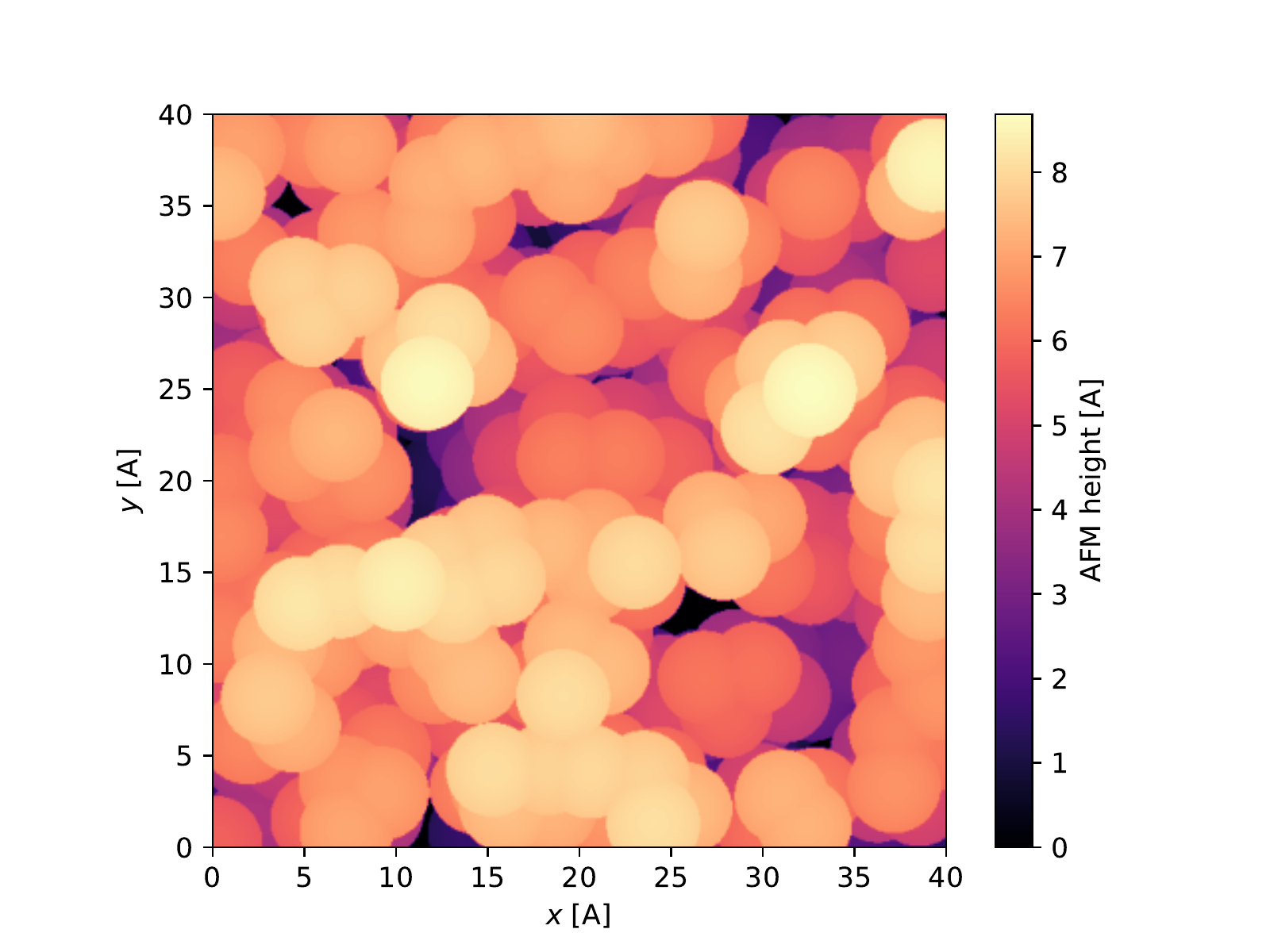} % Teflon is nr.1
    \caption{Simulated atomic force microscopy (AFM) images of Kapton (left) and Teflon (right). The width of the virtual AFM tip is 2.56 \AA.}
    \label{fig:MD2}
\end{figure}

To characterize the generated surfaces, we employed the technique by Włoch et al.\cite{Wloch2018} and simulated atomic force microscopy (AFM) patterns. The resulting AFM images are shown in Fig. S\ref{fig:MD2}.

The resulting roughness was determined by the root-mean-squared (RMS) calculation of the AFM surface data. The RMS values of our structures are shown in Table S1. These values coincide with that determined by Włoch et al., thereby confirming that we had a realistic surface structure.

\begin{table}[]
\caption{\label{tabS2} Root mean square roughness as determined from the AFM images.}
\begin{tabular}{lllll}
\cline{1-3}
\multicolumn{1}{|l|}{}  & \multicolumn{1}{l|}{Material} & \multicolumn{1}{l|}{RMS (\AA)} &  &  \\ \cline{1-3}
\multicolumn{1}{|l|}{1} & \multicolumn{1}{l|}{PTFE}  & \multicolumn{1}{l|}{3.76 $\pm$ 0.19} &  &  \\ \cline{1-3}
\multicolumn{1}{|l|}{2} & \multicolumn{1}{l|}{Kapton}  & \multicolumn{1}{l|}{3.98 $\pm$ 0.19} &  &  \\ \cline{1-3} & & &  & 
\end{tabular}
\end{table}

\begin{figure}[h]
    \centering
    \includegraphics[width=.2\textwidth]{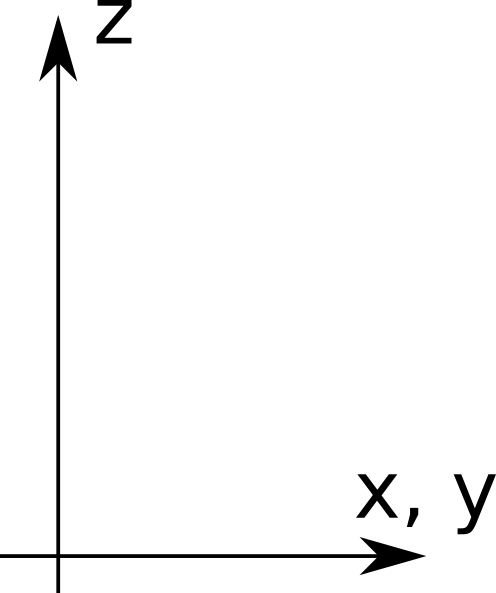}  % Axis illustration}
    \includegraphics[width=0.3\textwidth, trim=4.5cm 0 4.5cm 1cm, clip]{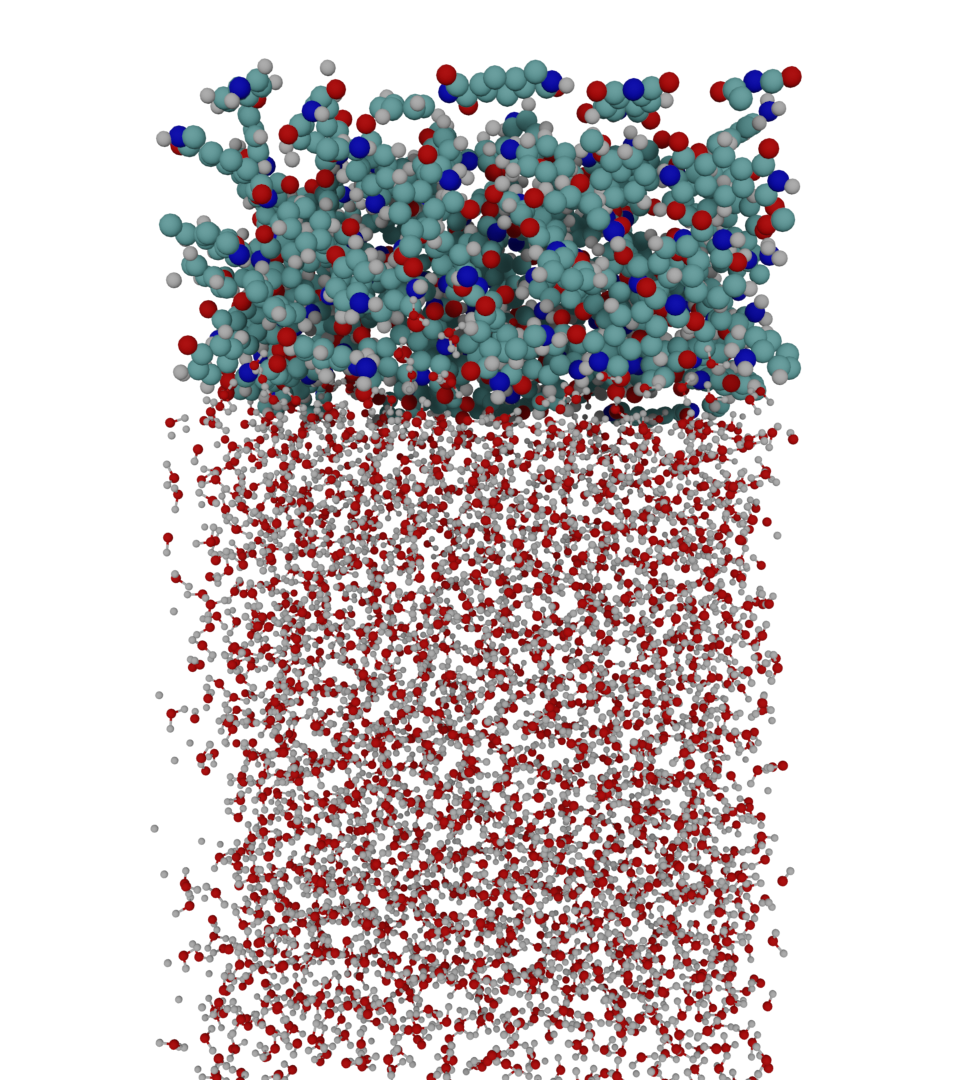}
    \includegraphics[width=.3\textwidth, trim=4.5cm 0 4.5cm 1cm, clip]{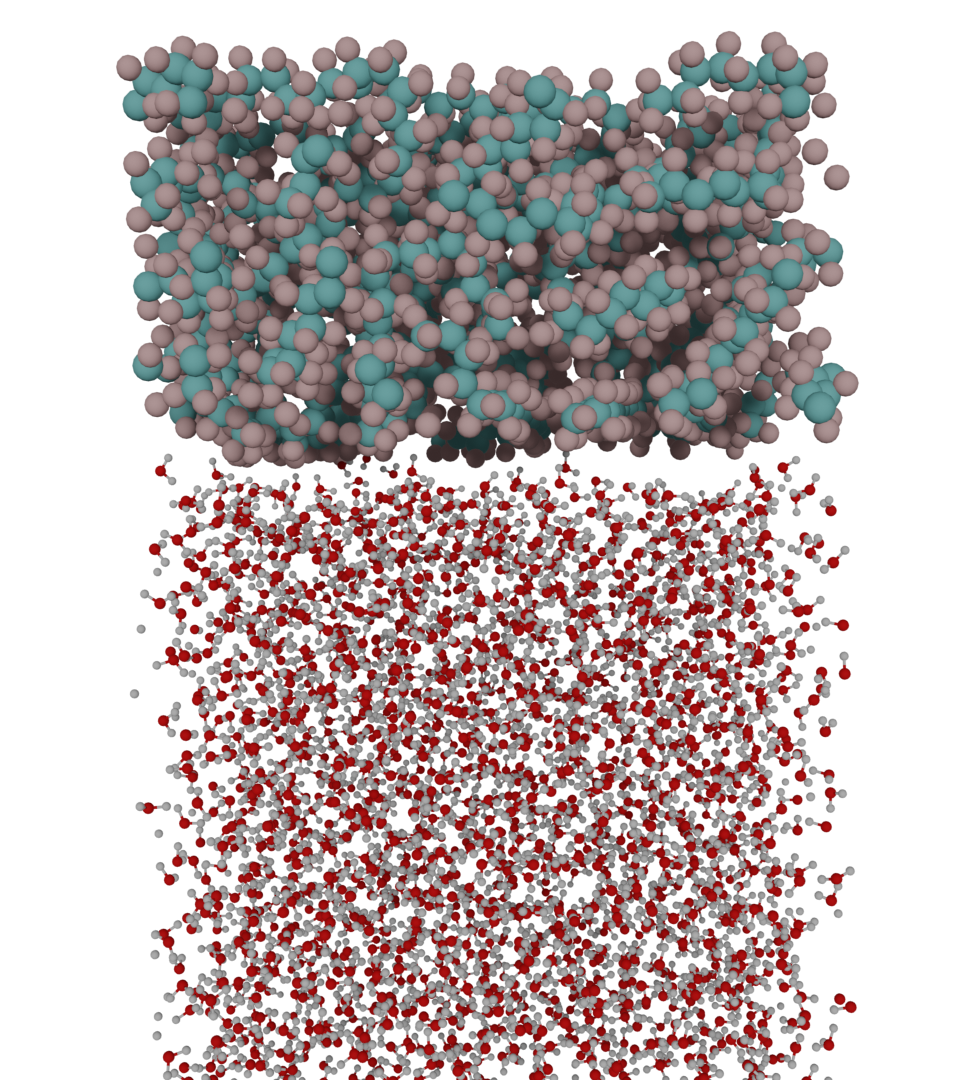}
    \caption{Partial simulation snapshot of the Kapton-water (left) and Teflon-water (right) systems. The system is periodic in the $x, y$-directions and limited by the wall structure in the $z$-direction. The difference in the hydrophobic/hydrophilic nature is also visible in the way that water gets much closer to the Kapton surface compared to that of Teflon.}
    \label{fig:MD3}
\end{figure}

To generate a capillary, we duplicate this structure, putting one on the far end in the $z$-direction and inserting a water slab of 10 nm. For all simulations, water is described by the SPC/E model~\cite{Berendsen1987}. For the hydronium (H$_3$O$^+$) and hydroxyl (OH$^-$) ions, we employ the parameters optimized by Bonthuis et al.~\cite{Bonthuis2016}. The free energy calculations were then performed after equilibration. Snapshots of both simulation systems are shown in Fig.~\ref{fig:MD3}. 

\subsection{2.2 Technical details of the free energy calculations}

We used free energy calculations to investigate the difference between the adsorption of ionic species on Kapton and Teflon surfaces because the relatively small concentrations of hydronium and hydroxyl ions compared to water molecules make system sizes too large for other methods. Specifically, we calculated the potential of mean force (PMF) for a single hydroxyl/hydronium ion. To be able to determine the PMF profiles, we utilize the umbrella sampling method~\cite{Kastner2011}, where the reaction coordinate of choice (in this case a line perpendicular to the capillary walls) is sampled in discrete windows. To let the particle explore a region around each window, a harmonic potential with a spring constant of k = 200 kcal/mol was used. By using the weighted histogram analysis method (WHAM), multiple windows along the reaction coordinate were combined and corrected for the additional biasing harmonic potential~\cite{Kumar1992, Souaille2001, Hub2010}. This process then finally results in a single, unified and unbiased free energy profile.

For each PMF, we initially used windows spaced about 0.5 nm apart from each other, which were run for a time of 2 ns at a time step of 2 fs. If the resulting histograms were not having sufficient overlap, we increased the number of windows. The system was equilibrated for 200 ps before each production run. To estimate the errors of the PMF profiles, we used bootstrapping analysis implemented by Hub et al~\cite{Hub2010}. In the plots shown in Fig. 3 (see main text), the error was estimated with 100 bootstrap samples. The plot had its reference point of $F = 0$ set to the mean value of the PMF in the region of 4 nm $<$ z $<$ 7 nm.

\begin{figure}[h]
    \centering
    \includegraphics[width=0.5\textwidth]{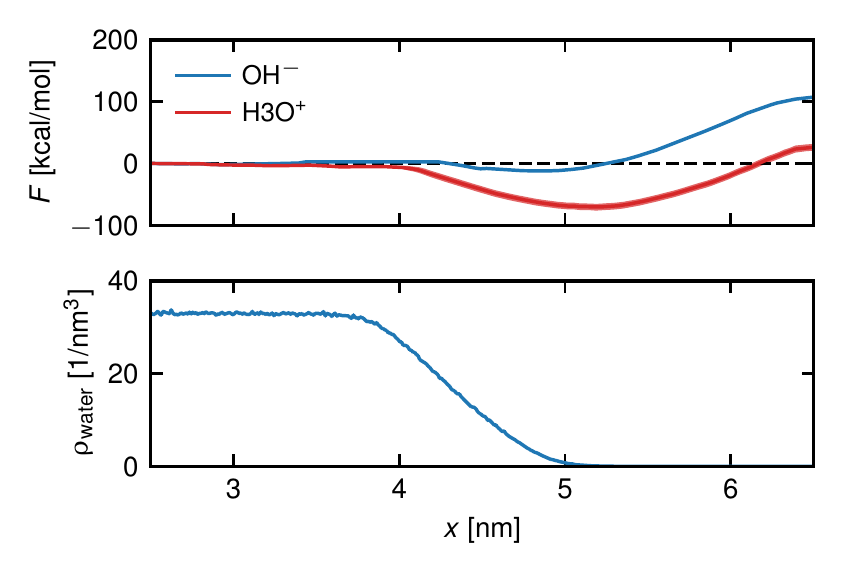}
    \caption{Potential of mean force of hydronium and hydroxyl ions at the water-vacuum interface (top panel), and the density of water molecules (bottom panel) near the interface region of the water droplet.}
    \label{fig:MD4}
\end{figure}

\begin{figure}[h]
    \centering
    \begin{minipage}[t]{.15\textwidth}
    \includegraphics[width=\linewidth]{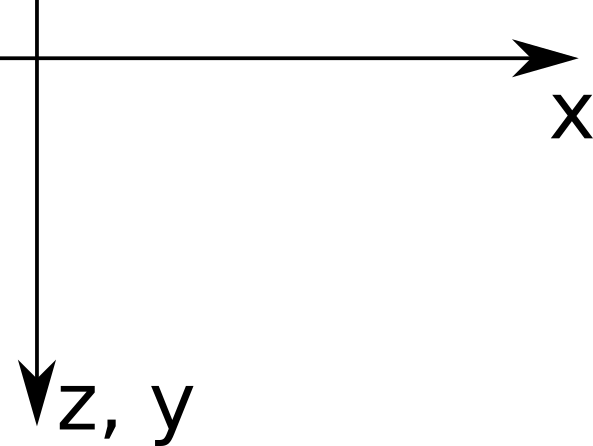}
    \end{minipage}
    \begin{minipage}[t]{.6\textwidth}
    \includegraphics[width=\linewidth]{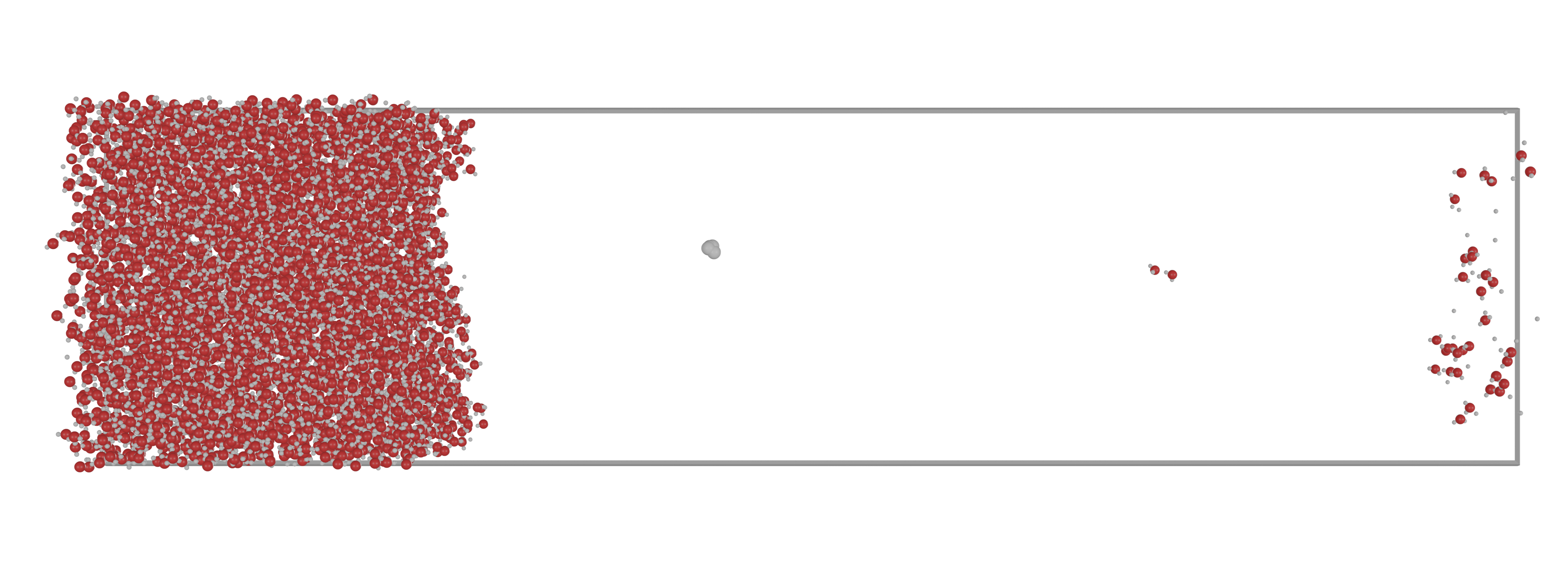}
    \end{minipage}
    \caption{Snapshot of the simulation system of a water-vapor interface. The fully periodic boundaries are illustrated by the grey border. During the simulations, the center of mass of the water slab is kept fixed by a spring potential to keep the water droplet from wandering through the box during sampling.}
    \label{fig:MD5}
\end{figure}

The adsorption of hydroxyl and hydronium ions on the droplet surface was also calculated for a vacuum-water interface by using PMF calculations (Fig. S\ref{fig:MD4}). The PMF exhibits very strong adsorption features of hydronium at the water-vapor interface (see a snapshot in Fig. S\ref{fig:MD5}). The hydronium ions exhibit stronger adsorption and extend further into the vacuum side of the interface, which is consistent with results obtained by Mamatkulov et al.~\cite{Mamatkulov2017}, where density profiles show a large peak of hydronium ions at the edge of the interface.

\begin{figure}[h]
    \centering
    \includegraphics[width=.49\textwidth]{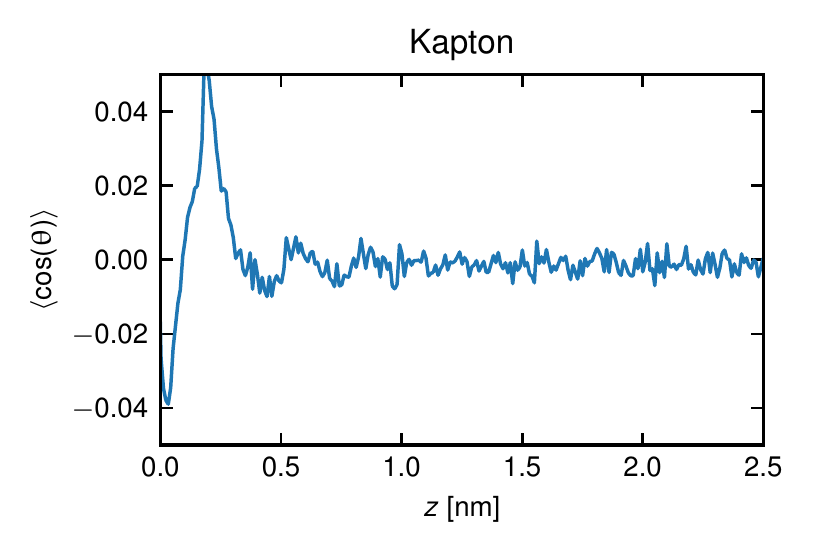}
    \includegraphics[width=.49\textwidth]{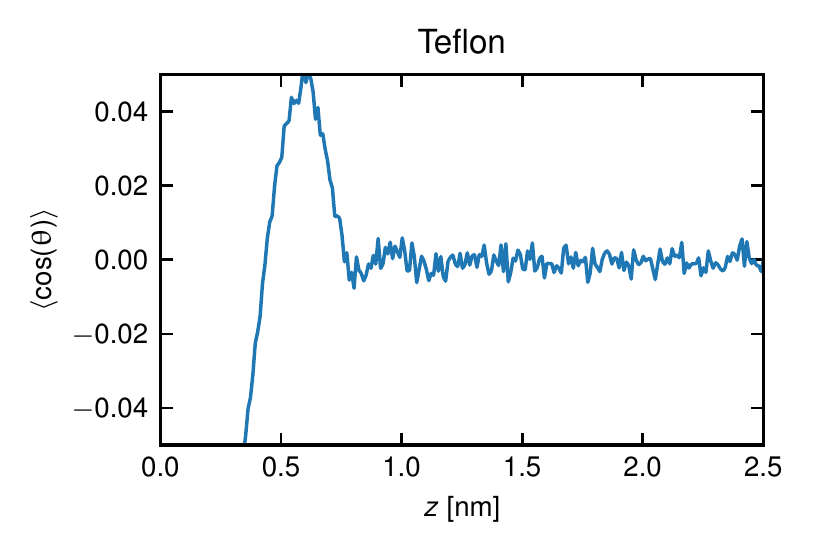}
    \caption{Orientation of water molecules near Kapton/Teflon surfaces. As in all plots, the location of the wall is given by $z= 0$.}
    \label{Fig:MD6}
\end{figure}

To see if there is a water molecules orientation near each surface, we calculated the orientation profile perpendicular to the wall, which is defined by:
$$
\rho _{\text{dipole}} = \cos( \theta(z))),
$$
where $\rho(z)$ is the number density profile and $\theta(z)$ is the angle of the molecular dipole with respect to the $z$-axis. The results of this analysis are shown in Fig. S\ref{Fig:MD6}. We observe no significant difference in the orientation away from the surface, but differences in water that is able to penetrate into the walls' surfaces. This leads us to the conclusion that a large part of the difference in the adsorption shown in the main text is not due to the surrounding medium of water molecules but due to interactions with the walls' surface.

\begin{figure}[h]
    \centering
    \includegraphics[width=.49\textwidth]{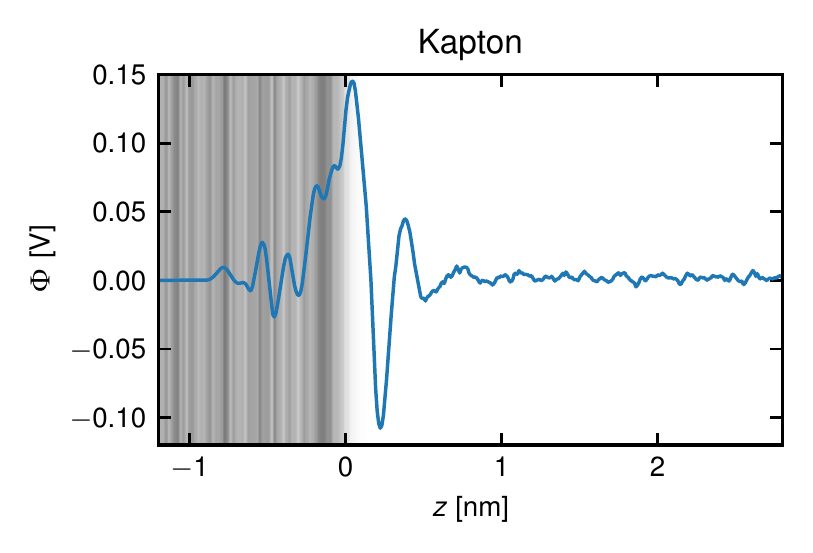}
    \includegraphics[width=.49\textwidth]{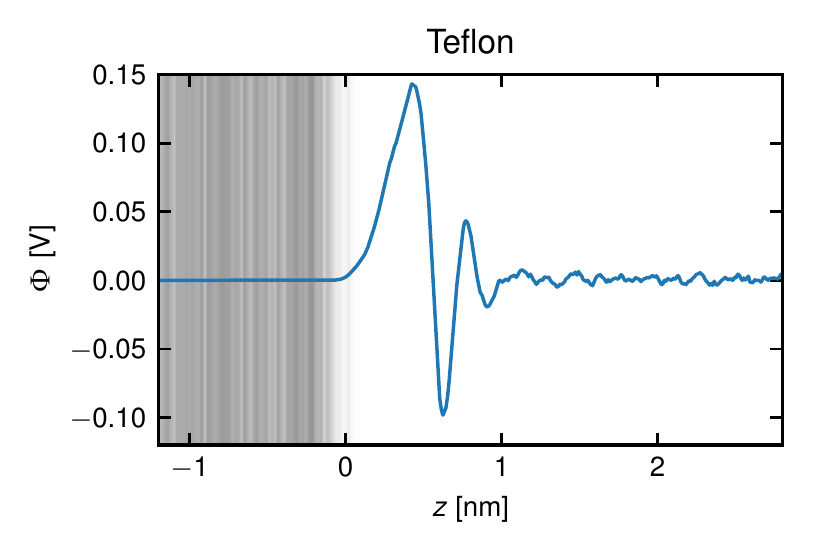}
    \caption{Electrostatic potential near Kapton/Teflon surfaces in a capillary filled with pure water.}
    \label{Fig:MD7}
\end{figure}

Figure S\ref{Fig:MD7} shows the calculated electrostatic potential near Kapton/Teflon surfaces in a capillary filled with water.
Similarly to the orientation profile, it does not show large differences between the different surface types.

\begin{figure}[h]
    \centering
    \includegraphics[width=\textwidth]{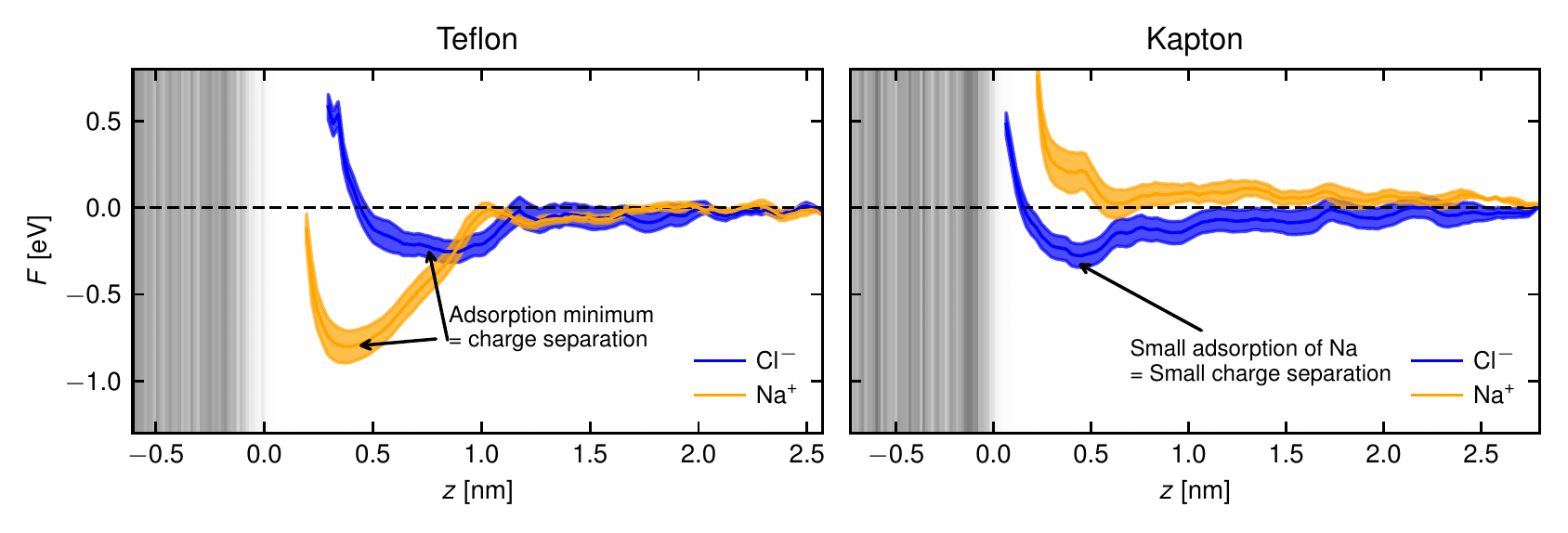}
    \caption{Potential of mean force of Na$^+$ and Cl$^-$ ions near the capillary surfaces.}
    \label{Fig:MD8}
\end{figure}

As the pH of the solution was varied by adding NaOH or HCl, we also investigated the affinity of Na$^+$ and Cl$^-$ ions at the capillary surfaces. The results of this investigation are shown in Fig. S\ref{Fig:MD8}, where we display the potential of mean force of both ion types for a Teflon surface on the left and for a Kapton surface on the right. We see similar results as the PMF shown in the main paper for hydronium and hydroxyl ions. The adsorption minima, especially of Na$^+$ ions, are significantly different when comparing Teflon to Kapton. Mirroring the results for H$_3$O$^+$ and OH$^-$ ions, we see different $z$-locations for the minima in Teflon, where Cl$^-$ has its adsorption minimum closer to the surface. Interestingly, there is a small absorption minimum for the Cl$^-$ ions near the Kapton surface, whereas Na$^+$ ions are only repelled by the surface and show no adsorption. Overall, these results mirror those shown in the main text and support the therein-described hypothesis for the difference in the droplet charge between Kapton and Teflon capillaries.

Note that to describe the boundary between solid and liquid phases, we used the notion of Gibbs dividing plane (or Gibbs surface). It is defined by $L_{G} = N_{w}/(A\cdot \rho_{\text{bulk}}) $, where $A$ is the surface area of the solid (capillary), $\rho_{\text{bulk}}$ is the bulk density of water molecules and $N_{w}$ is the number of water molecules. This surface can be understood as a location of a diffuse interface~\cite{Israelachvili2011}. The Gibbs plane shows how close the liquid interface is with respect to the solid (capillary) surface (see vertical dashed lines in Fig. 3, main text).

\bibliography{references_SI}

\section{3. Additional graphs, photos, and diagrams}

\begin{figure}
    \centering
    \includegraphics[scale=0.7]{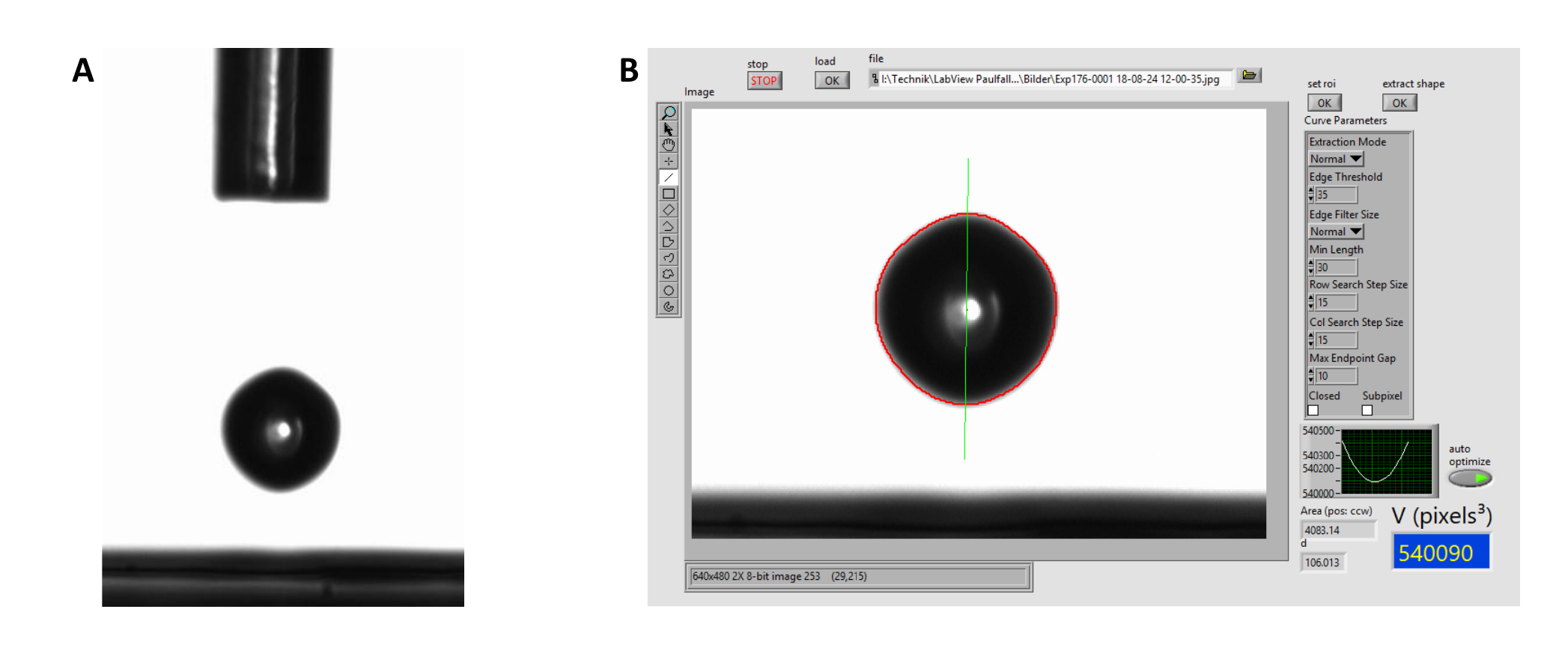}
    \caption{\textbf{Drop size determination:} (A) Photo of the drop before the Faraday cylinder. (B) Screenshot of the program for automatic volume determination by the drop shape (red line) and the axis of symmetry (green line).}
\label{fig:SI_1}
\end{figure}

\begin{figure}
    \centering
    \includegraphics[scale=0.30]{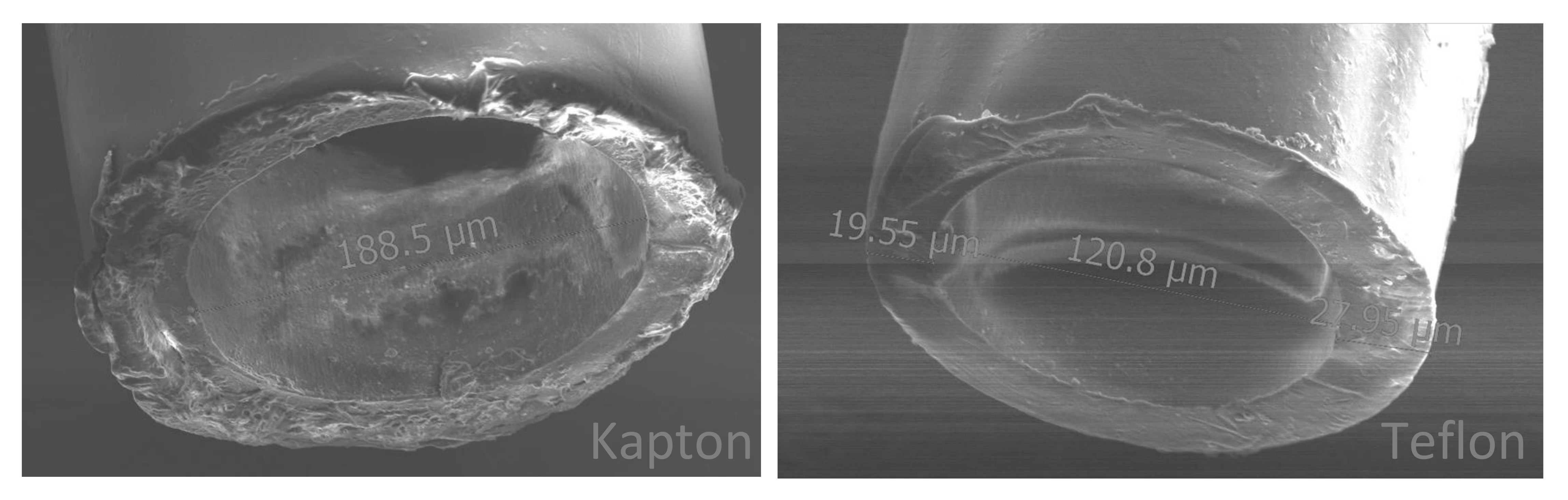}
    \caption{\textbf{Photos of the capillaries edges:} Kapton (left) and Teflon (right).}
\label{fig:SI_2}
\end{figure}

\begin{figure}
    \centering
    \includegraphics[scale=0.4]{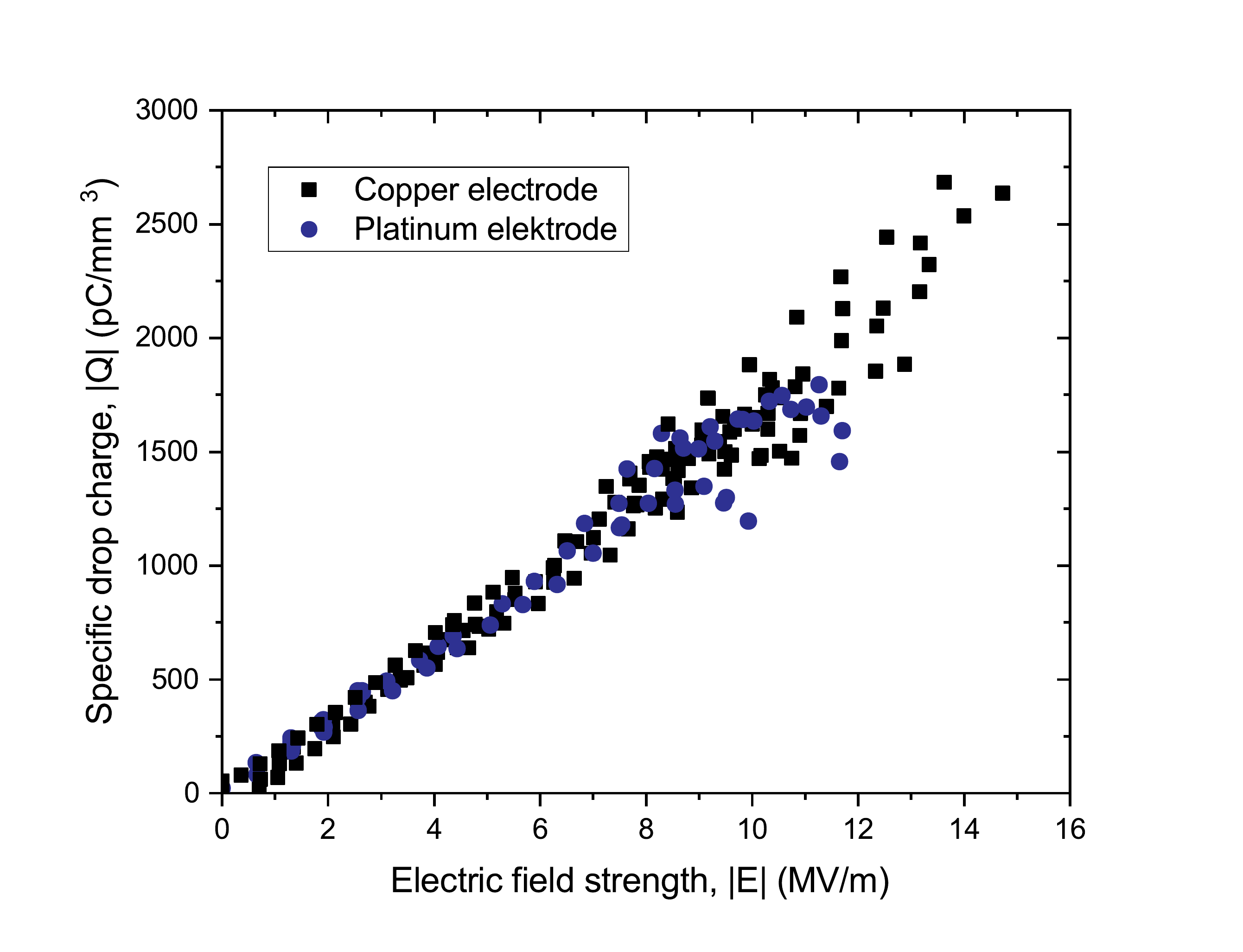}
    \caption{Dependence of the droplet charge on the electric field strength E for copper and platinum electrodes. No difference is observed in the whole E range.}
\label{fig:SI_3}
\end{figure}

\begin{figure}
    \centering
    \includegraphics[scale=0.93]{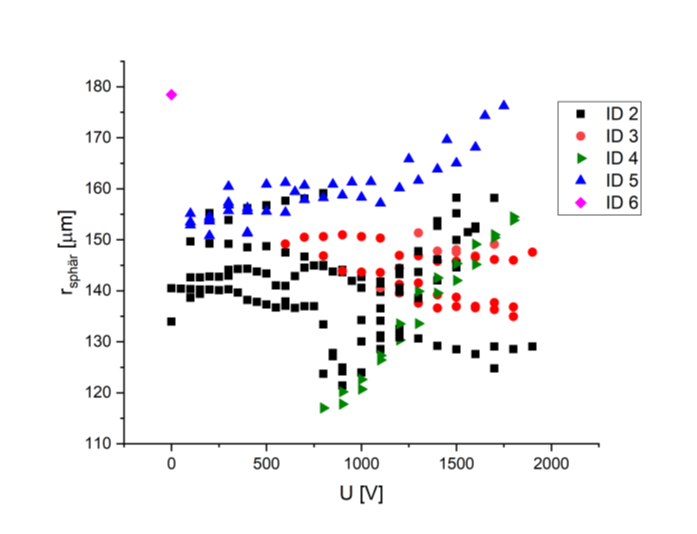}
    \caption{Radius of the drop vs voltage for different dispenser capillaries.}
\label{fig:SI_4}
\end{figure}

\begin{figure}
    \centering
    \includegraphics[scale=0.6]{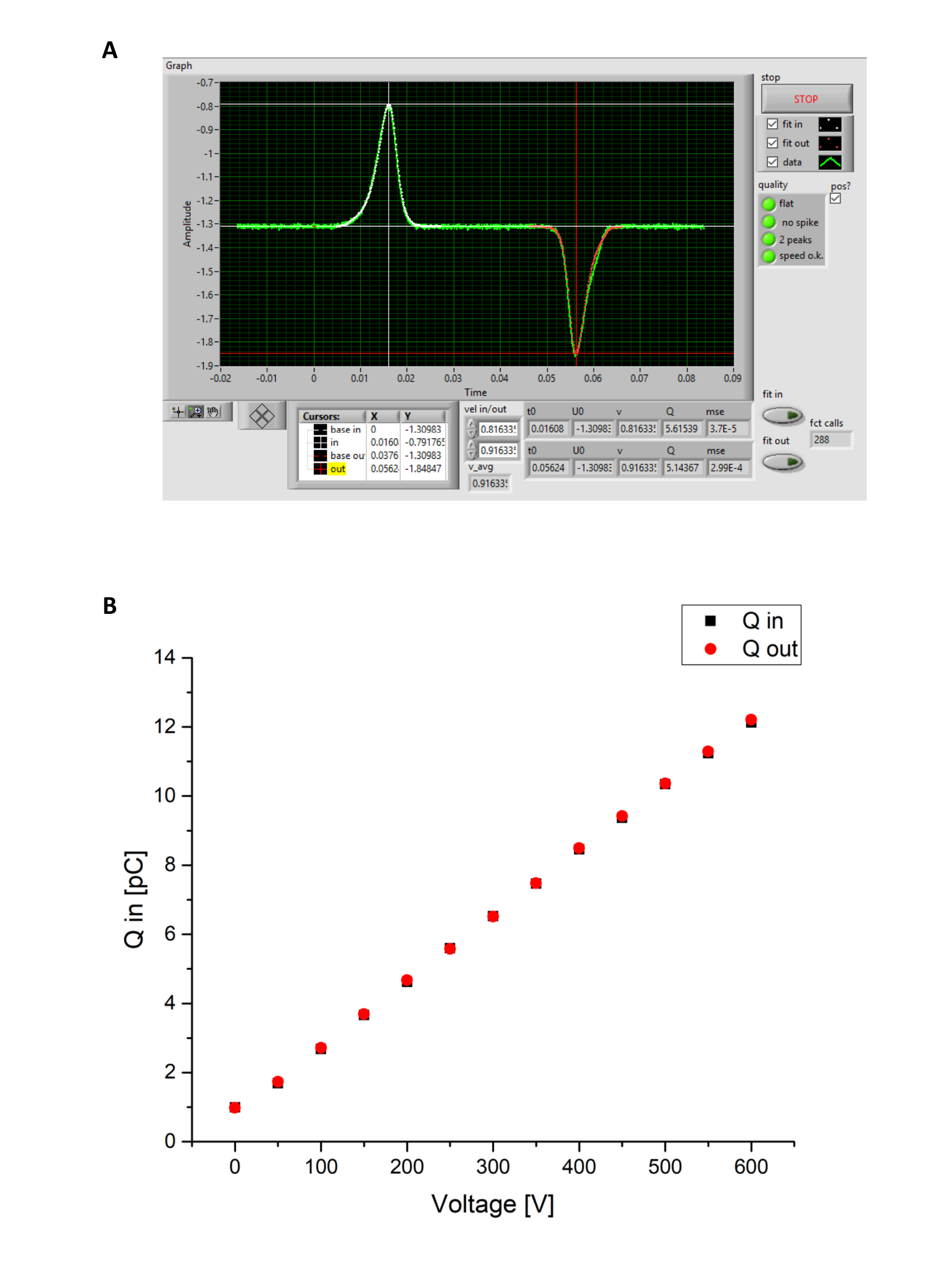}
    \caption{\textbf{Drop integrity test:} (A) Charge determination by integration of the positive (white) and the negative (red) parts of the FC signal. (B) Comparison of the charge obtained from the positive (black) and negative (red) semi-signals. If the curves do not coincide, the integrity of the drop is not guaranteed and such a result is excluded.
}
\label{fig:SI_5}
\end{figure}

\begin{figure}
    \centering
    \includegraphics[scale=0.7]{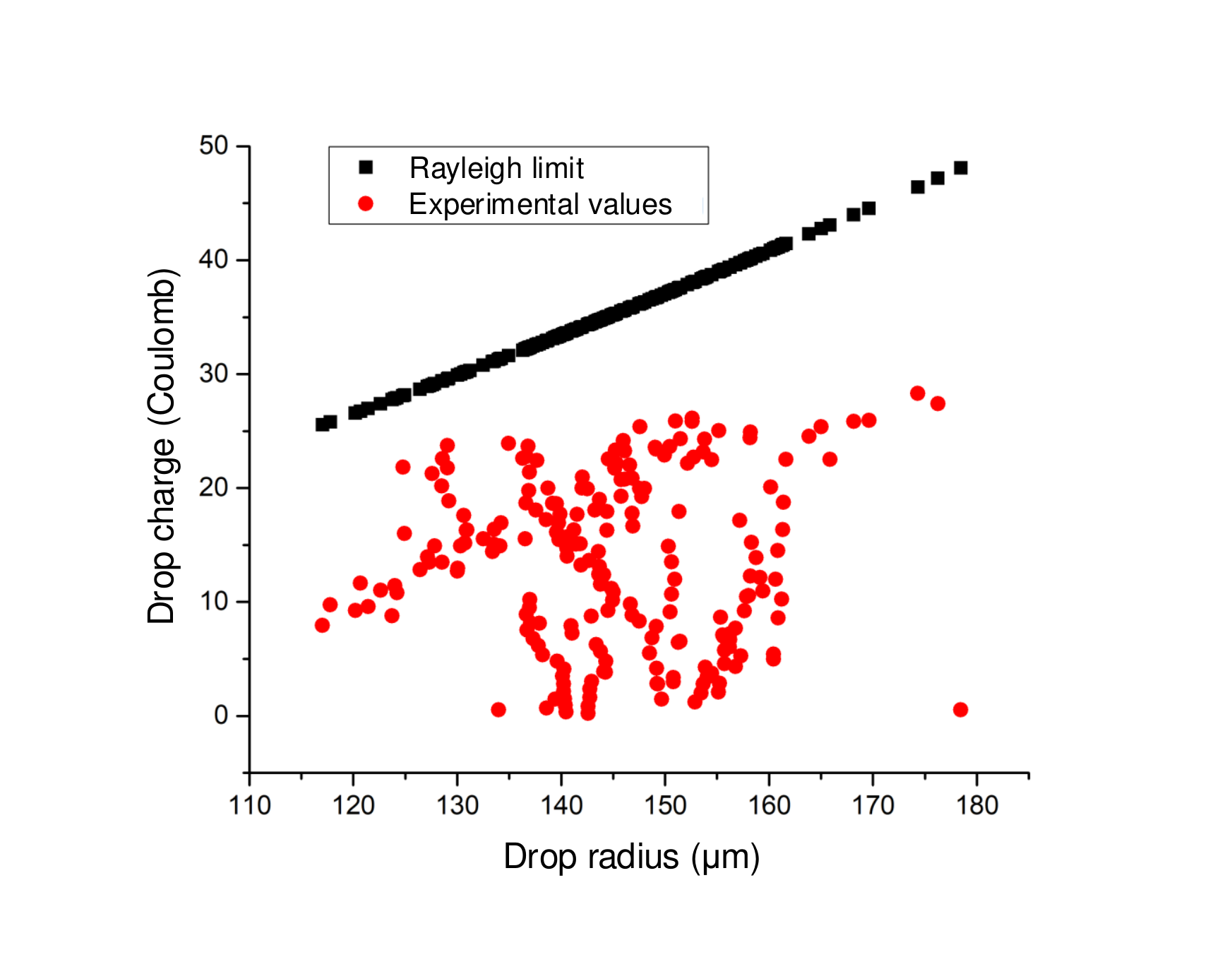}
    \caption{\textbf{Rayleigh stability test.} Charge vs radius of the drop for in experiment (red) and the theoretical maximum limit (black).}
\label{fig:SI_6}
\end{figure}

\begin{figure}
    \centering
    \includegraphics[scale=0.7]{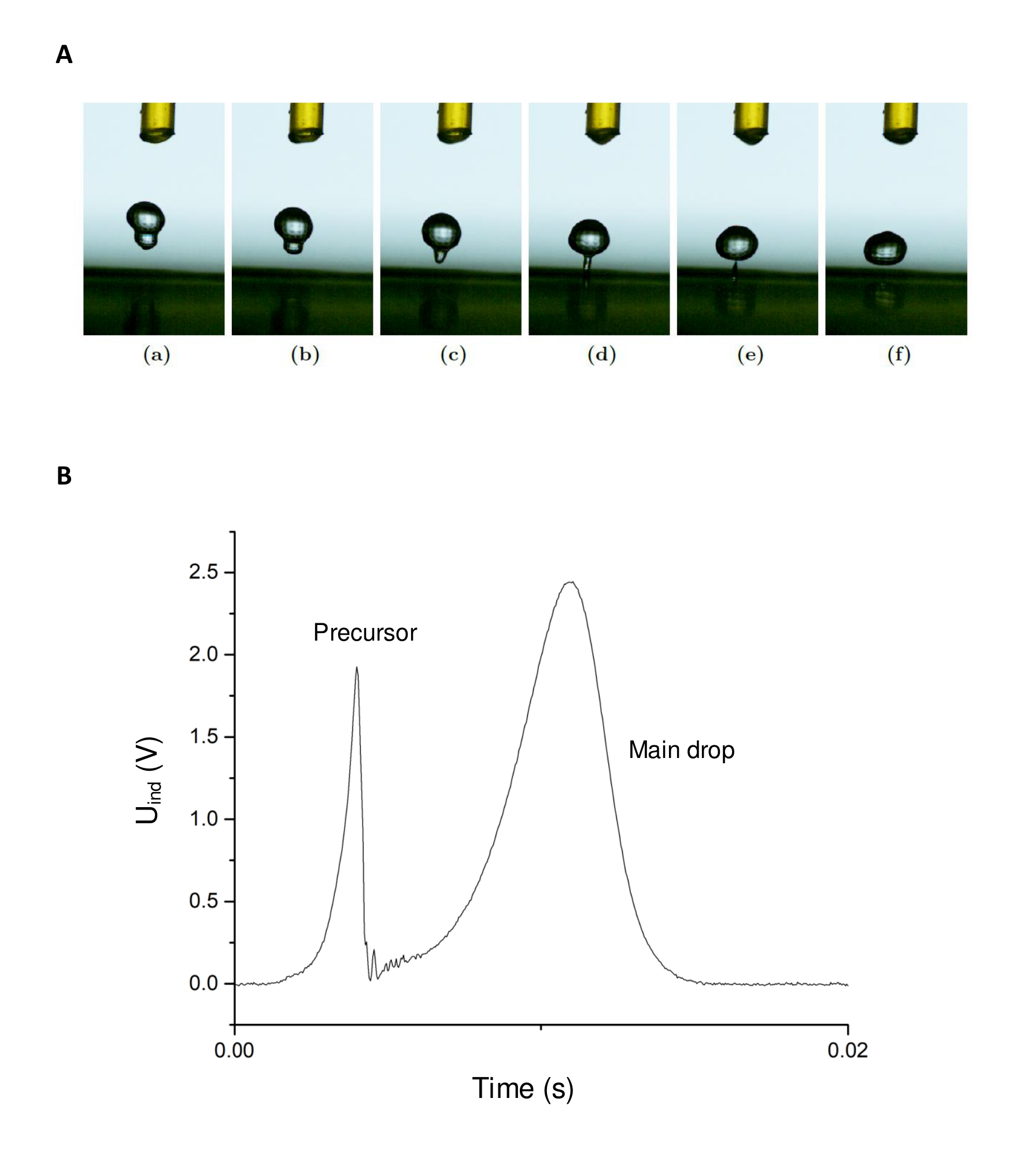}
    \caption{\textbf{Example of the possible drop instability:} (A) Series of the drop snapshots, showing a formation of precursor (a small drop accompanying the main drop) (B) Deviation of the FC signal due to the splitting of the drop due to Rayleigh instability. Inset shows the high-speed camera images for U=-1700 V. A splitting of the released drop can be seen shortly before the entrance of the FC.}
\label{fig:SI_7}
\end{figure}

\begin{figure}
    \centering
    \includegraphics[width=\linewidth]{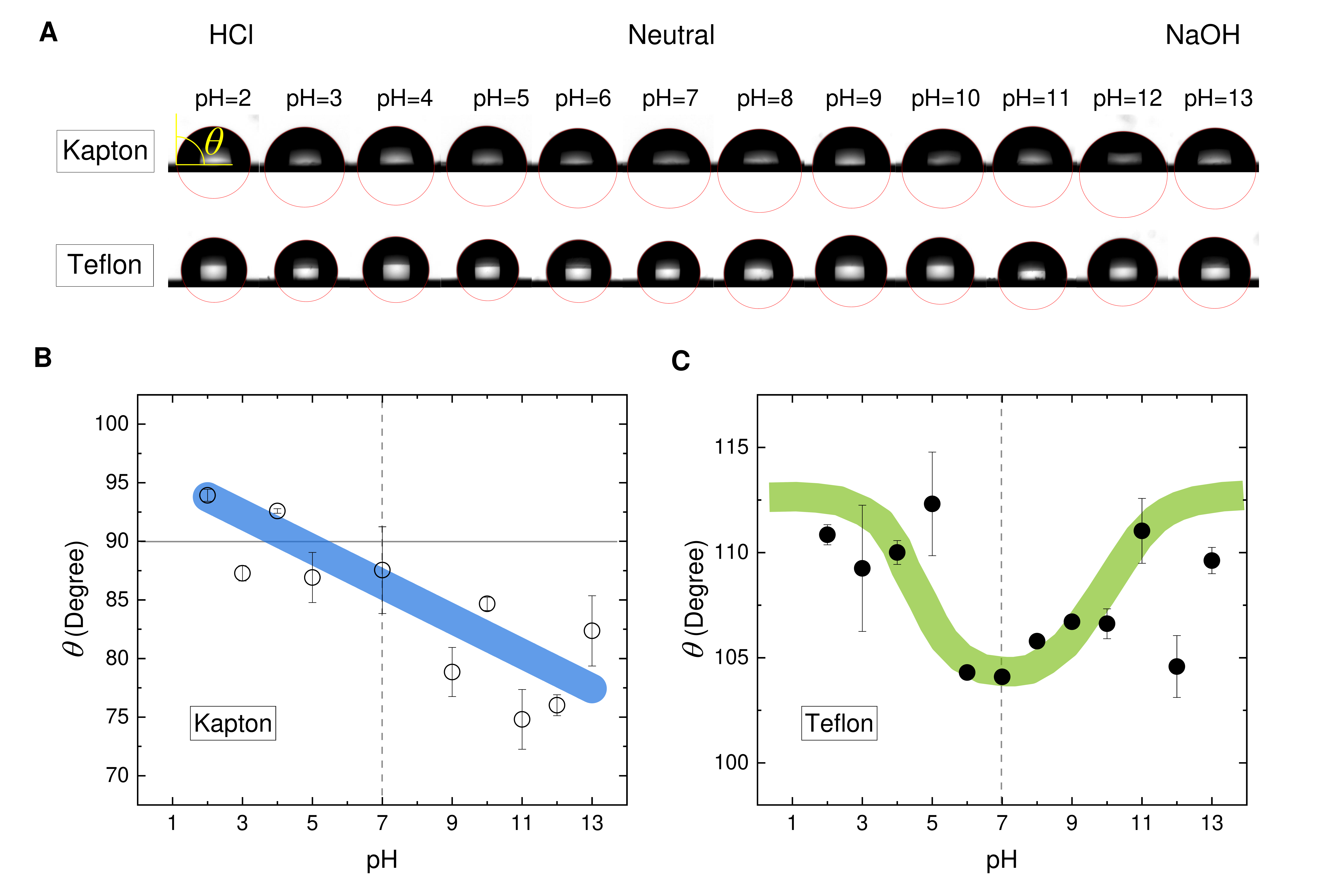}
    \caption{\textbf{Contact-angle vs pH.} (\textbf{A}) Photographs of the drops on Kapton and Teflon with different concentrations of HCl and NaOH. The red circles are guides for the eye. Note that although the circles look different, the actual amount of water in the drops is the same. (\textbf{B}) Dependencies of the contact angle on pH for Kapton.(\textbf{C}) The same for Kapton. The blue and green curves are guides for the eye.}
\label{fig:SI_8}
\end{figure}

\begin{figure}
    \centering
    \includegraphics[scale=0.4]{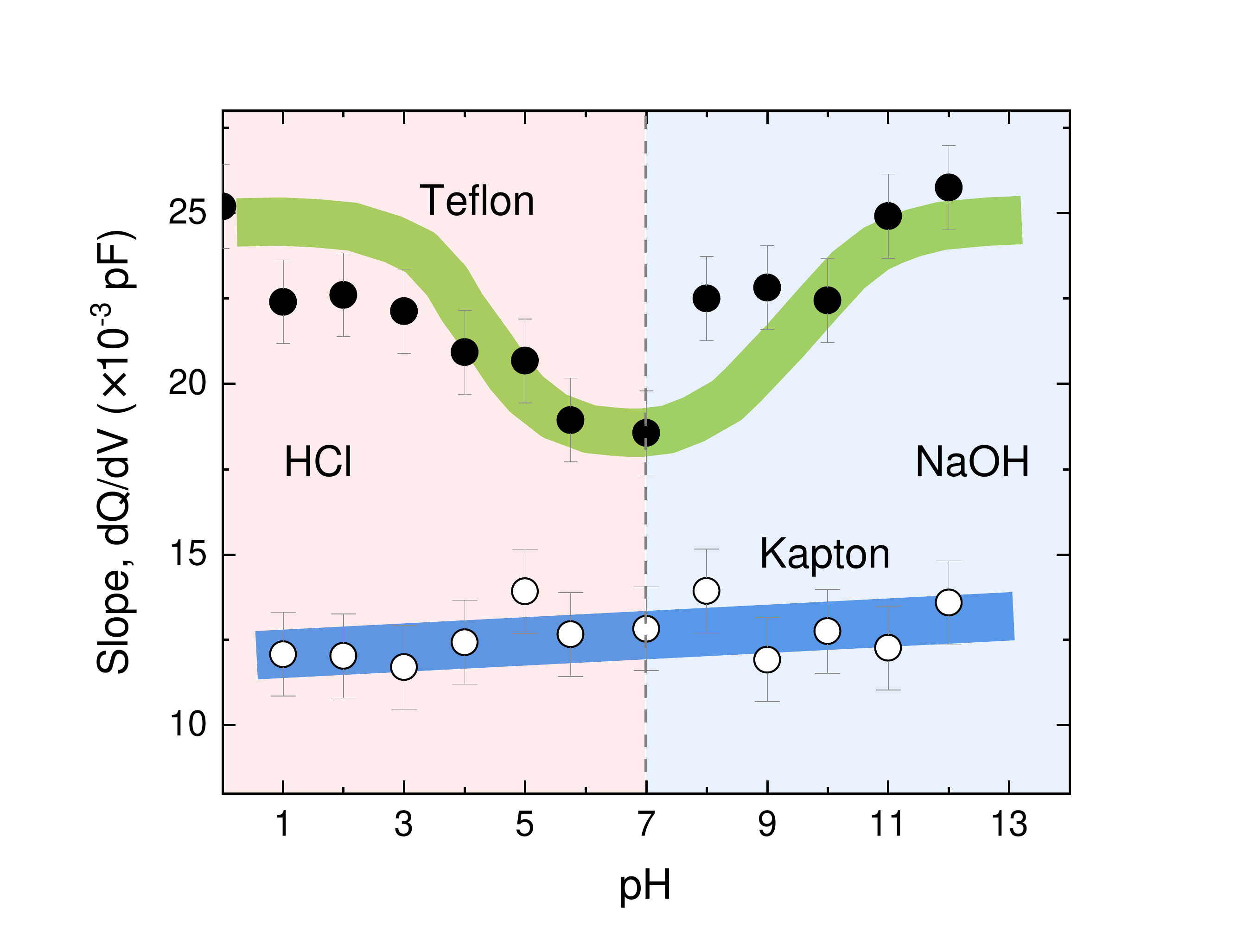}
    \caption{\textbf{Capacitance of the drops}. Dependence of the inclines of the curves from Fig. 1, D and E, (see main text) on pH. Kapton does not affect the drop capacitance, while Teflon shows an effect similar to that for the contact angle (Fig.S16)C.}
\label{fig:SI_9}
\end{figure}